\documentclass[prd,preprint,superscriptaddress,amsmath,amssymb]{revtex4}

\usepackage{graphicx,color,slashed}

\numberwithin{equation}{section}

\begin{document}

{\begin{flushright}{KIAS-P18039}
\end{flushright}}

\title{\bf \Large $\epsilon'_K/\epsilon_K$ and  $K \to \pi \nu \bar\nu$ in a two-Higgs doublet model}

\author{Chuan-Hung Chen}
\email{physchen@mail.ncku.edu.tw}
\affiliation{Department of Physics, National Cheng-Kung University, Tainan 70101, Taiwan}

\author{Takaaki Nomura}
\email{nomura@kias.re.kr}
\affiliation{School of Physics, KIAS, Seoul 02455, Korea}

\date{\today}

\begin{abstract}

The Kaon direct CP violation $Re(\epsilon'_K/\epsilon_K)$ below the experimental data of  a $2\sigma$ in the standard model, which is calculated using the  RBC-UKQCD lattice  and a large $N_c$ dual QCD,  indicates the necessity of a new physics effect. In order to resolve the insufficient $Re(\epsilon'_K/\epsilon_K)$, we study the charged-Higgs contributions  in a generic two-Higgs-doublet model. If we assume that the origin of the CP-violation phase is uniquely  from the Kobayashi-Maskawa (KM) phase when the constraints from the $B$- and $K$-meson mixings, $B\to X_s \gamma$, and Kaon indirect CP violating parameter $\epsilon_K$ are simultaneously taken into account, it is found that the Kaon direct CP violation through the charged-Higgs effects can reach $Re(\epsilon'_K/\epsilon_K)_{H^\pm}\sim 8 \times 10^{-4}$. Moreover, with the constrained values of the parameters, the branching ratios of the rare $K\to \pi \nu \bar\nu$ decays can be $BR(K^+\to \pi^+ \nu \bar\nu)\sim 13 \times 10^{-11}$ and $BR(K_L\to \pi^0 \nu \bar\nu)\sim 3.6 \times 10^{-11}$, where the results can be tested through the NA62 experiment at CERN and the KOTO experiment at J-PARC, respectively.

\end{abstract}

\maketitle

\section{Introduction}

Since a large time-dependent CP asymmetry through the $B_d \to J/\Psi K_S$ mode was observed at the BaBar~\cite{Aubert:2001sp} and BELLE~\cite{Abashian:2001pa} experiments, it is certain that the origin of the observed CP violation at colliders, including the indirect ($\epsilon_K$) and direct ($Re(\epsilon'_K/\epsilon_K)$) CP violation  in the $K$-meson,   mainly stems from the unique CP phase of the Cabibbo-Kobayashi-Maskawa (CKM) matrix~\cite{Cabibbo:1963yz,Kobayashi:1973fv} in the standard model (SM). 

Recently, the RBC-UKQCD collaboration reported  the surprising lattice QCD results on the matrix elements of $K\to \pi \pi$ and $Re(\epsilon'_K/\epsilon_K)$~\cite{Boyle:2012ys,Blum:2011ng,Blum:2012uk,Blum:2015ywa,Bai:2015nea}, where the Kaon direct CP violation and the contribution from the electroweak penguin to it are respectively given as~\cite{Blum:2015ywa,Bai:2015nea}:
 \begin{equation}
 Re(\epsilon'_K/\epsilon_K) = 1.38(5.15)(4.59) \times 10^{-4}\,, \quad  Re(\epsilon'_K/\epsilon_K)_{\rm EWP}= -(6.6 \pm 1.0) \times 10^{-4}\,,
 \end{equation}
whereas the average of the  NA48~\cite{Batley:2002gn} and KTeV~\cite{AlaviHarati:2002ye,Abouzaid:2010ny} results is $Re(\epsilon'_K/\epsilon_K)=(16.6 \pm 2.3)\times 10^{-4}$. That is,  a $2.1\sigma$ below the experimental value is obtained using the lattice calculations. 

Intriguingly, the recent theoretical calculations of $Re(\epsilon'_K/\epsilon_K)$ using a large $N_c$ dual QCD  approach~\cite{Buras:2015xba, Buras:2015yba}, which was developed by~\cite{Buras:1985yx,Bardeen:1986vp,Bardeen:1986uz,Bardeen:1986vz,Bardeen:1987vg}, support the RBC-UKQCD results, and the results are obtained as:
 \begin{align}
 Re(\epsilon'_K/\epsilon_K)_{\rm SM} &=  (8.6 \pm 3.2)\times 10^{-4}\,, ~~\text(B^{(1/2)}_6=B^{(3/2)}_8=1)\,, \\
  Re(\epsilon'_K/\epsilon_K)_{\rm SM} & =  (6.0 \pm 2.4) \times 10^{-4}\,, ~~ \text(B^{(1/2)}_6=B^{(3/2)}_8=0.76)\,,
 \end{align}
where $B^{(1/2)}_6$ and $B^{(3/2)}_8$ denote the non-perturbative parameters of gluon and electroweak penguin operators, respectively. It is found that both approaches obtain consistent values of $B^{(1/2)}_6$ and $B^{(3/2)}_8$ as~\cite{Buras:2015xba}:
 \begin{align}
& B^{(1/2)}_6 (m_c) = 0.57\pm 0.19  \,, \ B^{(3/2)}_8(m_c)=0.76 \pm 0.05 ~~  \text{(RBC-UKQCD)}\,, \nonumber \\
&  B^{(1/2)}_6 \leq B^{(3/2)} < 1 \,, \ B^{(3/2)}_8(m_c)=0.80 \pm 0.1\,. ~~~~~~~~~~~ \text{(large $N_c$)}\,.
 \end{align}
Since the main contributions to $Re(\epsilon'_K/\epsilon_K)$ in the SM are dictated by  $B^{(1/2)}_{6}$ and $B^{(3/2)}_8$, and there is a  cancellation between gluon and electroweak penguin contribution, thus,  a  smaller $B^{(1/2)}_6$  leads to a $Re(\epsilon'_K/\epsilon_K)_{\rm SM}$ below the experimental value of $2\sigma$.  The small $Re(\epsilon'_K/\epsilon_K)$, which results from a QCD based approach and arises from  the short-distance (SD) effects, could be compensated for by other sources in the SM, such as chromomagnetic dipole effects and  long-distance (LD) final state interactions (FSIs). However, according to the recent study in~\cite{Buras:2018evv}, the contribution  to $Re(\epsilon'_K/\epsilon_K)$ from the gluonic dipole operators in the SM should be less than $10^{-4}$ and cannot explain the data. In addition, the conclusion about the LD contribution is still uncertain, where the authors in~\cite{Buras:2016fys} obtained  negative result, but the authors in~\cite{Gisbert:2017vvj} obtained $Re(\epsilon'_K/\epsilon_K)=(15\pm7) \times 10^{-4}$ when the SD and LD effects were considered. Hence, in spite of the  large uncertainty of the current lattice calculations, if we take the RBC-UKQCD's central result, which basically includes all nonperturbative QCD effects, as the tendency of the SM,  the alternative source for the insufficient $Re(\epsilon'_K/\epsilon_K)$ can be attributed to  a new physics effect~\cite{Buras:2015qea,Buras:2015yca,Buras:2015kwd,Buras:2015jaq,Tanimoto:2016yfy,Buras:2016dxz,Kitahara:2016otd,Endo:2016aws,Bobeth:2016llm,Cirigliano:2016yhc,Endo:2016tnu,Bobeth:2017xry,Crivellin:2017gks,Bobeth:2017ecx,Haba:2018byj}, which we will focus on in this study.

In rare $K$ decays,  two important  unobserved processes are $K^+ \to \pi^+ \nu \bar\nu$ and $K_L \to \pi^0 \nu \bar\nu$, where the former is a CP-conserving channel, and the latter  denotes a CP-violation. The NA62 experiment at CERN plans to measure the branching ratio (BR) for $K^+ \to \pi^+ \nu \bar\nu$, which can reach the SM result with  a $10\%$ precision~\cite{Rinella:2014wfa,Moulson:2016dul},  and the KOTO experiment at J-PARC  will observe the $K_L \to \pi^0 \nu \bar\nu$ decay~\cite{Komatsubara:2012pn,Beckford:2017gsf}. In addition to their sensitivity to new physics, their importance is that the SM predictions are  theoretically clean, where the QCD corrections at the next-to-leading-order (NLO)~\cite{Buchalla:1993bv,Misiak:1999yg,Buchalla:1998ba} and NNLO~\cite{Gorbahn:2004my,Buras:2005gr,Buras:2006gb} and the electroweak corrections at the NLO~\cite{Buchalla:1997kz,Brod:2008ss,Brod:2010hi} have been calculated. 
The SM predictions are~\cite{Buras:2015qea}:
 \begin{align}
 BR(K^+\to \pi^+ \nu \bar \nu) &= (9.11 \pm 0.72) \times 10^{-11} \,, \\
 BR(K_L \to \pi^0 \nu\bar \nu) & = (3.00\pm 0.31)\times 10^{-11}\,,
 \end{align}
whereas the current experimental situations are $BR(K^+\to \pi^+ \nu \bar \nu)^{\rm exp} =(17.3^{+11.5}_{-10.5})\times 10^{-11}$~\cite{Artamonov:2008qb} and $BR(K_L \to \pi^0 \nu\bar \nu)^{\rm exp} < 2.6\times 10^{-8}$~\cite{Ahn:2009gb}. Recently, NA62 reported its first result using the 2016 taken data. It was found that  one candidate event of $K^+\to \pi^+ \nu \bar \nu$ is observed and that the upper bound of BR is given  by $BR(K^+\to \pi^+ \nu \bar\nu)< 14 \times 10^{-10}$ at a $95\%$ confidence level (CL)~\cite{Pinzino:2018pmz}. 

To pursue  new physics contributions to the $Re(\epsilon'_K/\epsilon_K)$ and rare $K$ decays, in this work, we investigate the influence of a charged-Higgs in a generic two-Higgs-doublet model (2HDM), i.e., the type-III 2HDM, where  global symmetry is not imposed on the Yukawa sector. As a result, flavor changing neutral currents (FCNCs) in such models can arise at the tree level. To reasonably suppress the tree-level FCNCs for the purpose of satisfying the constraints from the $B$ and $K$ systems, such as $\Delta M_{B_d, B_s}$, $B\to X_s \gamma$, $\Delta M_K$, and $\epsilon_K$, we can adopt the so-called Cheng-Sher ansatz~\cite{Cheng:1987rs}, where the neutral scalar-mediated flavor-changing effects  are dictated by the square-root of the mass product of the involved flavors, denoted by $\sqrt{m_{f_i} m_{f_j}}/v$. Thus, we can avoid extreme fine-tuning of the free parameters when they contribute to the rare $K$ and $B$ decays. The alternative approach for suppressing the FCNCs using 't Hooft's naturalness criterion~\cite{tHooft:1979rat} can be found in~\cite{Crivellin:2013wna}, where more related flavor phenomena were studied in detail. 

From a phenomenological viewpoint, the  reasons why the charged-Higgs effects in the type-III 2HDM are interesting can be summarized as follows: firstly, $Re(\epsilon'_K/\epsilon_K)$ and $K\to \pi \nu \bar\nu$ in the SM are all dictated by the product of the CKM matrix elements $V^*_{ts}$ and $V_{td}$. The same CKM factor automatically appears in the charged-Higgs Yukawa couplings without introducing any new weak CP-violation phase; thus, we can avoid the strict limits from the time-dependent CP asymmetries in the $B_d$ and $B_s$ systems. Secondly, unlike the type-II 2HDM, where the  charged-Higgs mass is bounded to be $m_{H^\pm}> 580$ GeV via the $B\to X_s \gamma$ decay~\cite{Misiak:2017bgg,Chen:2018hqy}, the charged-Higgs in the type-III model can be much lighter than that in the type-II model, due to the modification of the Yukawa couplings~\cite{Chen:2018hqy}. Thirdly, a peculiar unsuppressed Yukawa coupling $\sqrt{m_c/m_t}V_{cq'}/V_{tq'}\chi^{u*}_{ct}$ (see the later discussions), which originates from the FCNCs,  also appears in the charged-Higgs couplings to the top-quark and down-type quarks~\cite{Chen:2018hqy,Chen:2017eby}. The effects play a key role in enhancing $Re(\epsilon'_K/\epsilon_K)$ and $K\to \pi \nu \bar\nu$ in this model.  Fourthly, the charged-Higgs effects can naturally provide the  lepton-flavor universality violation and can be used to resolve the puzzles in the semileptonic $B$ decays, such as $R(D)$, $R(K^{(*)})$, and large $BR(B^-_u \to \tau \bar\nu)$~\cite{Chen:2018hqy,Chen:2017eby,Dutta:2013qaa,Bhattacharya:2014wla,Bhattacharya:2016mcc,Dutta:2016eml,Bhattacharya:2016zcw,Dutta:2017xmj,Akeroyd:2017mhr,Arhrib:2017yby}. 

Since the charged-Higgs effects have a strong correlation with different phenomena, the new free parameters are not constrained by only one physical observable. Therefore, the involved new parameters are strictly limited and cannot be arbitrarily free. It is found that when the constraints of $\Delta B=2$, $\Delta K=2$, $B\to X_s \gamma$, and $\epsilon_K$ are simultaneously taken into account, the charged-Higgs contribution to the direct CP violation of $K$-meson can reach $Re(\epsilon'_K/\epsilon_K)_{H^\pm}\sim 8 \times 10^{-4}$ (not including the SM contribution), and the BR for $K^+ \to \pi^+ \nu \bar\nu$ can be  $BR(K^+ \to \pi^+ \nu \bar\nu)\sim 13 \times 10^{-11}$, while $BR(K_L \to \pi^0 \nu \bar\nu)\sim 3.6 \times 10^{-11}$.

The paper is organized as follows: In Section II, we briefly review the charged-Higgs and neutral scalar Yukawa couplings to the fermions with the Cheng-Sher ansatz in the type-III 2HDM. In Section III, we formulate $\Delta M_K$, $\epsilon_K$, and $Re(\epsilon'_K/\epsilon_K)$ in the 2HDM. The charged-Higgs contributions to the rare $K\to \pi \nu \bar\nu$ decays are shown in Section IV. The detailed numerical analysis is shown in Section V, where  the constraints from $\Delta M_{B_d, B_s}$, $B\to X_s \gamma$, $\Delta M_K$, and $\epsilon_K$ are included. A conclusion is given in Section V.  

\section{ Charged and neutral Higgs couplings to the quarks and leptons}

In this section, we summarize the Yukawa couplings of the neutral Higgses and charged-Higgs to the quarks and leptons in the generic 2HDM. The Yukawa couplings without imposing extra global symmetry can be in general written as:
\begin{align}
-{\cal L}_Y &=  \bar Q_L Y^d_1 D_R H_1 + \bar Q_L Y^{d}_2 D_R H_2
+ \bar Q_L Y^u_1 U_R \tilde H_1 + \bar Q_L Y^{u}_2 U_R \tilde H_2
 \nonumber \\
&+  \bar L Y^\ell_1 \ell_R H_1 + \bar L Y^{\ell}_2 \ell_R H_2 + H.c.\,, 
\label{eq:Yu}
\end{align}
where all flavor indices are hidden; $P_{R(L)}=(1\pm \gamma_5)/2$; $Q_L$ and $L_L$ are the $SU(2)_L$ quark and lepton doublets, respectively; $f_R$ ($f=U,D,\ell$) denotes the singlet  fermion;  $Y^f_{1,2}$ are the $3\times 3$ Yukawa matrices, and $\tilde H_i = i\tau_2 H^*_i$. There are two CP-even scalars, one CP-odd pseudoscalar, and two charged-Higgs particles in the 2HDM, and the relations between physical and weak states can be expressed as: 
\begin{align}
 h &= -s_\alpha \phi_1 + c_\alpha  \phi_2 \,, \nonumber \\
 H&= c_\alpha \phi_1 + s_\alpha \phi_2 \,, \nonumber \\
H^\pm (A) &= -s_\beta \phi^\pm_1 (\eta_1) + c_\beta \phi^\pm_2 (\eta_2)\,, \label{eq:hHHA}
\end{align}
where $\phi_i(\eta_i)$ and $\eta^\pm_i$ denote the real (imaginary) parts of the neutral and charged components  of $H_i$, respectively;  $c_\alpha (s_\alpha)= \cos\alpha (\sin\alpha)$, $c_\beta  = \cos\beta = v_1/v$, and $s_\beta= \sin\beta = v_2/v$,  $v_{i}$ are the vacuum expectation values (VEVs) of $H_i$, and $v=\sqrt{v^2_1 + v^2_2}\approx 246$ GeV. In this study, $h$ is the SM-like Higgs while $H$, $A$, and $H^\pm$ are new particles in the 2HDM. 

Introducing the unitary matrices $V^f_{L,R}$ to diagonalize the quark and lepton mass matrices, the Yukawa couplings of scalars $H$ and $A$ can then be obtained as:
 \begin{align}
-{\cal L}^{H, A}_{Y} &= \bar u_L \left[ \frac{s_\alpha}{v s_\beta} {\bf m_u} + \frac{s_{\beta\alpha}}{s_\beta} {\bf X}^u \right] u_R H 
 + \bar d_L \left[ \frac{c_\alpha}{v c_\beta} {\bf m_d} - \frac{s_{\beta\alpha}}{ c_\beta} {\bf X}^d \right] d_R H \nonumber \\
&+ \bar \ell_L \left[ \frac{c_\alpha}{v c_\beta} {\bf m_\ell} - \frac{s_{\beta\alpha}}{ c_\beta} {\bf X}^\ell \right] \ell_R H 
+ i  \bar u_L \left[ - \frac{c_\beta}{v } {\bf m_u} + \frac{{\bf X}^u}{ s_\beta}  \right] u_R A  \nonumber \\
&+ i \bar d_L \left[ -\frac{t_\beta}{v } {\bf m_d} + \frac{{\bf X}^d}{ c_\beta}  \right] d_R A
+ i \bar \ell_L \left[ -\frac{t_\beta}{v } {\bf m_\ell} + \frac{{\bf X}^\ell }{c_\beta}  \right] \ell_R A + H.c.\,, \label{eq:HAY}
\end{align}
where ${\bf m_{f}}$ is the diagonalized fermion mass matrix; $t_\beta=s_\beta/c_\beta=v_2/v_1$; $c_{\beta\alpha} = \cos(\beta-\alpha)$; $s_{\beta\alpha}= \sin(\beta-\alpha)$, and ${\bf X}^fs$ are defined as:
 \begin{align}
 {\bf X^u}= V^u_L \frac{Y^u_1}{\sqrt{2}} V^{u\dagger}_R \,, \     {\bf X^d}= V^d_L \frac{Y^d_2}{\sqrt{2}} V^{d\dagger}_R \,, \ 
{\bf X^\ell}= V^\ell_L \frac{Y^\ell_2}{\sqrt{2}} V^{\ell\dagger}_R\,. \label{eq:Xfs}
\end{align}
We can also obtain the Higgs Yukawa couplings; however, it is found that the  associated $X^{f}$ terms are always related  to $c_{\alpha\beta}$, which is strictly bound by the current precision Higgs data. For simplicity, we take the alignment limit with $c_{\alpha\beta}=0$ in the following analysis. Thus, the Higgs couplings are the same as those in the SM. 
The  charged-Higgs Yukawa couplings to fermions are found as:
\begin{align}
-{\cal L}^{H^\pm}_Y &=  \sqrt{2} \bar d_L V^\dagger  \left[ - \frac{ 1}{v t_\beta } {\bf m_u} + \frac{{\bf X}^u}{ s_\beta}  \right] u_R H^- \nonumber \\ 
&+ \sqrt{2} \bar u_L V \left[ - \frac{ t_\beta}{v } {\bf m_d} + \frac{{\bf X}^d}{  c_\beta}  \right] d_R H^+  \nonumber \\
&+\sqrt{2} \bar \nu_L  \left[ -\frac{t_\beta}{v } {\bf m_\ell} + \frac{{\bf X}^\ell }{ c_\beta}  \right] \ell_R H^+ + H.c.\,, \label{eq:CHY}
\end{align}
where $V\equiv V^u_L V^{d^\dagger}_L$ stands for the CKM matrix. Except the factor $\sqrt{2}$ and  CKM matrix, the Yukawa couplings of charged Higgs are the same as those of pseudoscalar $A$. 

 From  Eq.~(\ref{eq:HAY}), the FCNCs at the tree level can be induced through the ${\bf X^f}$ terms. To suppress the tree-induced $\Delta F=2$ ($F=K,B_{d(s)},D$) processes,  we employ the Cheng-Sher ansatz~\cite{Cheng:1987rs}  as:
\begin{equation}
X^f_{ij} =\frac{ \sqrt{m_{f_i} m_{f_j}}}{v} \chi^f_{ij}\,, \label{eq:CSA}
 \end{equation}
 where  $\chi^f_{ij}$ are the new free parameters. With the Cheng-Sher ansatz,  the Yukawa couplings of scalars $H$ and $A$ to the down-type quarks can be straightforwardly  obtained as:
 \begin{align}
-{\cal L}^{H, A}_{Y} &=   \frac{t_\beta}{v} \bar d_{i L} \left[  m_{d_i} \delta_{ij} - \frac{\sqrt{m_{d_i} m_{d_j}}}{ s_\beta} \chi^d_{ij} \right] d_{j R} (H -i A) + H.c. \,,\label{eq:Yu_HA}
\end{align}
where the CKM matrix elements are not involved. 

Since the charged-Higgs interactions are associated with the CKM matrix elements, the couplings involving the third generation quarks may not be small; therefore, for the $K$-meson decays, it is worth  analyzing  the charged-Higgs Yukawa couplings  of the $d$- and $s$-quark to the top-quark, i.e., $tdH^+$ and $tsH^+$.  According to Eq.~(\ref{eq:CHY}), the  $t_R d_L H^+$ coupling can be written and simplified as
 \begin{align}
 t_R d_L H^+  & : \frac{\sqrt{2}}{v} \left[ \left( \frac{1}{t_\beta} - \frac{\chi^{u*}_{tt} }{s_\beta}\right) m_t V_{td} - \frac{\sqrt{m_t m_c}}{s_\beta} \chi^{u*}_{ct} V_{cd} - \frac{\sqrt{m_t m_u}}{s_\beta} \chi^{u*}_{ut} V_{ud}\right] \nonumber \\
 & \approx  \sqrt{2} \frac{m_t}{v } V_{td} \left( \frac{1}{t_\beta} - \frac{\chi^L_{td}}{s_\beta}  \right)\,, \quad \chi^L_{td} = \chi^{u*}_{tt} +\sqrt{\frac{m_c}{m_t}} \frac{V_{cd}}{V_{td}} \chi^{u*}_{ct}\,, \label{eq:CLtd}
 \end{align}
where  we have dropped the $\chi^{u*}_{ut}$ term because its coefficient is a factor of 4 smaller than the $\chi^{u*}_{ct}$ term.  In addition to the $m_t$ enhancement,  the effect associated with $\chi^u_{ct}$ is $\sqrt{m_c/m_t} |V_{cd}/V_{td}| \chi^{u*}_{ct} \approx 2.4 \chi^{u*}_{ct}$, which  is in principle not suppressed.  Intriguingly, the charged-Higgs coupling is comparable to the SM gauge coupling of $(g/\sqrt{2})V_{td}$. Because $m_d V_{td} \ll \sqrt{m_s m_d} V_{ts}  \ll \sqrt{m_b m_d} V_{tb}$,  the  $t_L d_R H^+$ coupling can be approximated  as:
 \begin{equation}
t_L d_R H^+ : - \sqrt{2} \frac{m_b t_\beta }{v }  \sqrt{\frac{ m_d}{m_b}}\frac{\chi^d_{bd} V_{tb}}{s_\beta}\,. \label{eq:CRtd}
 \end{equation}
 Although there is no $V_{td}$ suppression, because  $\chi^{d}_{bd} \sim O(10^{-2})$ is constrained by the $B_d$ mixing~\cite{Chen:2018hqy}, the coupling of $t_L d_R H^+$ is somewhat smaller than that of $t_R d_L H^+$, even with the large value of $t_\beta$, e.g. $t_\beta\sim 50$. Using $m_t |V_{ts}| \sim 6.72 \ {\rm GeV} < \sqrt{m_c m_t} V_{cs} \sim 14.8$ GeV and   $ m_s V_{ts} \ll \sqrt{m_s m_b} V_{tb} \sim 0.66$ GeV, the $t s H^+$ coupling can be similarly obtained as:
  \begin{align}
t_R s_L H^+ & :  \sqrt{2} \frac{m_t}{v} V_{ts} \left( \frac{1}{t_\beta} - \frac{\chi^L_{ts}}{s_\beta}\right)\,, \quad \chi^L_{ts} = \chi^{u*}_{tt} + \sqrt{\frac{m_c}{m_t}} \frac{V_{cs}}{V_{ts}} \chi^{u*}_{ct}\,, \label{eq:CLts} \\
 t_L s_R H^+  & :  -\sqrt{2}\frac{m_b t_\beta}{v } \sqrt{\frac{m_s}{ m_b}} \frac{\chi^d_{bs} V_{tb}}{s_\beta}\,. \nonumber 
  \end{align}
The detailed analysis for the other charged-Higgs couplings can be found in~\cite{Chen:2018hqy}. In sum, the charged-Higgs couplings to the $d(s)$- and top-quark in the type-III 2HDM can be formulated as:
 \begin{align}
 {\cal L}^{H^\pm}_Y & \supset  \frac{\sqrt{2}}{v} V_{tq'} \bar t  \left( m_t \zeta^u_{tq'} P_L  -  m_b \zeta^d_{tq'} P_R\right) q'  H^+ +H.c.,  \label{eq:tqH}
 \end{align}
where  the parameters $\zeta^{f}_{ij}$ are defined as:
\begin{align}
\zeta^u_{tq'} & = \frac{1}{t_\beta} - \frac{\chi^L_{tq'} }{s_\beta}\,, \ \chi^L_{tq'}  =\chi^{u*}_{tt} + \sqrt{\frac{m_c}{m_t}} \frac{V_{cq'}}{V_{tq'}} \chi^{u*}_{ct}\,, \nonumber \\
\zeta^{d}_{tq'} & = t_\beta  \sqrt{\frac{m_{q'}}{m_b}} \frac{V_{tb}}{V_{tq'}}  \frac{\chi^d_{bq'}}{s_\beta} \,.
  \label{eq:zetas}
\end{align}
For the lepton sector,  we use  the flavor-conserving scheme with ${\bf X}^\ell_{ij} = (m_{\ell_i} /v )  \chi^\ell_{\ell_i} \delta_{\ell_i \ell_j}$, i.e. $\chi^\ell_{\ell_i \ell_j} = \chi^\ell_{\ell_i}  \delta_{\ell_i \ell_j}$; as a result, the Yukawa couplings of $H^\pm$ to the leptons can be expressed as:
  \begin{equation}
  {\cal L}^{H^\pm}_{Y,\ell} = \sqrt{2} \frac{\tan\beta \, m_\ell}{v}  \zeta^\ell_\ell \bar \nu_{\ell }  P_R \ell  H^+ + H.c.\,, \label{eq:Clepton}
  \end{equation}
 with  $\zeta^\ell_{\ell}=1 - \chi^{\ell}_{\ell}/s_\beta$. The suppression factor $m_\ell /v$ could be moderated using a large value of  $\tan\beta$. In this work, we use the interactions shown in Eqs.~(\ref{eq:Yu_HA}), (\ref{eq:tqH}), and (\ref{eq:Clepton}) to study the influence on the $K^0-\bar K^0$ mixing $\Delta M_K$, $\epsilon_K$, $\epsilon'_K/\epsilon_K$, and $K\to \pi \nu \bar \nu$ decays.

\section{ Formulations of $\Delta M_K$, $\epsilon_K$, and $\epsilon'_K/\epsilon_K$ in the generic 2HDM}

\subsection{$\Delta M_K$ and $\epsilon_K$}

To study the new physics contributions to $\Delta M_K$ and $\epsilon_K$, we follow the notations in~\cite{Buras:2001ra} and write  the effective Hamiltonian for $\Delta S=2$  as:
 \begin{equation}
 {\cal H}^{\Delta S=2} = \frac{ G^2_F}{16 \pi^2} m^2_W \sum_i   V^i_{\rm CKM} C_{i} (\mu) Q_{i} \,, 
 \end{equation}
where $V^i_{\rm CKM}$ are the involved CKM matrix elements; $C_i(\mu)$ are the Wilson coefficients at the $\mu$ scale,  and the relevant operators $Q_i$ are given as:
 \begin{align}
 Q^{VLL}_1 & = (\bar s^\alpha \gamma_\mu P_L d^\alpha) (\bar s^\beta \gamma^\mu P_L d^\beta)\,, \nonumber \\
 Q^{LR}_1 & = (\bar s^\alpha \gamma_\mu P_L d^\alpha) (\bar s^\beta \gamma^\mu P_R d^\beta)\,, \nonumber \\
 Q^{LR}_2 & = (\bar s^\alpha P_L d^\alpha) (\bar s^\beta P_R d^\beta)\,, \nonumber \\
 Q^{SLL}_1 & = (\bar s^\alpha P_L d^\alpha) (\bar s^\beta P_L d^\beta)\,, \nonumber \\
 Q^{SLL}_{2} & = (\bar s^\alpha \sigma_{\mu\nu} P_L d^\alpha) (\bar s^\beta \sigma^{\mu \nu} P_L d^\beta). \label{eq:Qs}
 \end{align}
The operators $Q^{VRR}_1$ and $Q^{SRR}_{i}$ can be obtained from $Q^{VLL}_1$ and $Q^{SLL}_{i}$ by switching $P_R$ and $P_L$, respectively.  

   %%%%
\begin{figure}[phtb]
\includegraphics[scale=0.7]{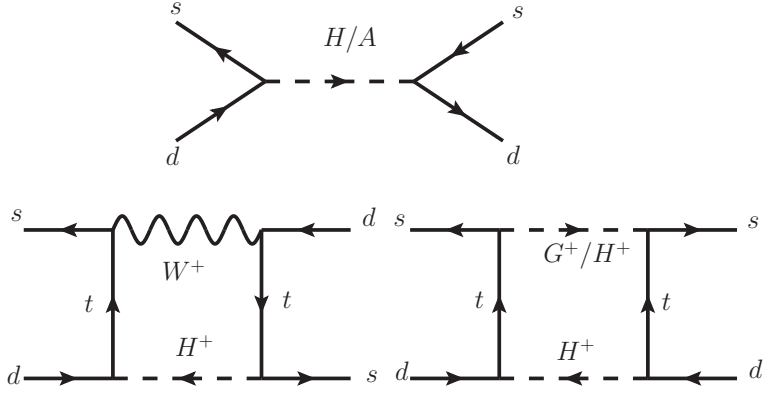}
 \caption{Sketched Feynman diagrams for the $\Delta S=2$  process.  }
\label{fig:DS_2}
\end{figure}

In the type-III 2HDM, the $\Delta S=2$ process can arise from the $H/A$-mediated tree  FCNCs and the $H^\pm$-mediated box diagrams, for which the representative Feynman diagrams are sketched in Fig.~\ref{fig:DS_2}. According to the interactions in Eq.~(\ref{eq:Yu_HA}), the $H/A$-induced effective Hamiltonian can be expressed as:
 \begin{align}
 {\cal H}^{\Delta S=2}_S & = -\frac{G^2_F}{16\pi^2 } m^2_W  \left[ C^{SLL}_{S1} Q^{SLL}_1 + 
 C^{SRR}_{S1} Q^{SRR}_1 + C^{LR}_{S2} Q^{LR}_2 \right]\,, \label{eq:HS2}
 \end{align}
where the subscript $S$ denotes the scalar and pseudoscalar contributions. Clearly,  no CKM matrix elements are involved in the tree FCNCs. Since the involved operators are $Q^{SLL,SRR}_1$ and $Q^{LR}_2$,  the corresponding Wilson coefficients at the $\mu_S$ scale are obtained as:
\begin{align}
C^{SLL}_{S1} & = r_S \left( \frac{1}{m^2_H} - \frac{1}{m^2_A}\right) ( \chi^{d*}_{ds} )^2  \,, \nonumber \\
C^{SRR}_{S1} & =  r_S \left( \frac{1}{m^2_H} - \frac{1}{m^2_A}\right) ( \chi^{d}_{sd} )^2 \,, \nonumber \\
C^{LR}_{S2} & = r_S \left( \frac{1}{m^2_H} +  \frac{1}{m^2_A}\right) 2 \chi^d_{sd} \chi^{d*}_{ds} \,, \nonumber \\
r_S & = \frac{4 \sqrt{2} \pi^2 \tan^2\beta \sqrt{x_d x_s}}{ G_F  s_\beta}\approx 1.0 \times 10^{3} \left( \frac{\tan^2\beta}{50^2 s_\beta} \right)\,, \label{eq:WCSK2}
\end{align}
with $x_q = m^2_q/m^2_W$. It can be seen from $r_S$  that although the $H/A$ effects are suppressed by $m_d m_s/m^2_W$, due to the $\tan\beta$ enhancement, the $\Delta M_K$ through the intermediates of $H$ and $A$ becomes sizable. We then can use the  measured $\Delta M_K$ to  bound  the parameters $\chi^d_{sd}$ and $\chi^d_{ds}$. If we take $m_H= m_A$ and $\chi^d_{sd}= \chi^d_{ds}$, it can be seen that $C^{SLL}_{S1}=C^{SRR}_{S1}=0$ and that  $C_{S2}^{LR} \propto \chi_{sd}^d \chi^{d*}_{ds} = |\chi^d_{sd}|^2$ is a real parameter; under this condition, $\epsilon_K$ that has arisen from the neutral scalars will be suppressed. 

From the charged-Higgs interactions  in Eq.~(\ref{eq:tqH}), we find that  with the exception of $Q^{SLL}_2$, the $W^\pm$-$H^\pm$, $G^\pm$-$H^\pm$, and $H^\pm$-$H^\pm$ box diagrams can induce all operators shown in Eq.~(\ref{eq:Qs}); the associated CKM matrix element factor is $V^i_{\rm CKM} =(V^*_{ts} V_{td})^2$, and  the Wilson coefficients at the $\mu=m_{H^\pm}$ scale can be expressed as:
\begin{align}
C^{VLL}_{H^\pm 1} & = 4 \left( 2 y_t^2 I^{WH}_1 (y_t, y_W)+ x_t y_t I^{WH}_{2} (y_t, y_W)\right) \zeta^{u*}_{ts} \zeta^u_{td} + 2 x_t y_t \left( \zeta^{u*}_{ts} \zeta^{u}_{td} \right)^2 I^{HH}_1(y_t) \,, \nonumber \\
C^{VRR}_{H^\pm 1} & = 2 x_b y_b  (\zeta^{d*}_{ts} \zeta^d_{td})^2 I^{HH}_1 (y_t) \,, \nonumber \\ 
C^{LR}_{H^\pm 1} & = 4 x_b y_t (\zeta^{u*}_{ts} \zeta^u_{td}) (\zeta^{d*}_{ts} \zeta^d_{td}) I^{HH}_1(y_t)\,, \nonumber \\
C^{LR}_{H^\pm 2} & = -8 \frac{m^2_b}{m^2_t} (\zeta^{d*}_{ts} \zeta^{d}_{td}) \left(  x_t y^2_t I^{WH}_1(y_t, y_W) + 2 y_t I^{WH}_2 (y_t, y_W) \right) \nonumber \\
& -8 x_b y^2_t (\zeta^{u*}_{ts} \zeta^{d}_{td}) (\zeta^{d*}_{ts} \zeta^{u}_{td}) I^{HH}_2 (y_t)\,, \nonumber \\ 
C^{SLL}_{H^\pm 1}  & = - 4 x_b y^2_t (\zeta^{d*}_{ts} \zeta^{u}_{td})^2 I^{HH}_{2} (y_t) \,, \nonumber \\
C^{SRR}_{H^\pm 1}  & = -4 x_b y^2_t (\zeta^{u*}_{ts} \zeta^{d}_{td})^2 I^{HH}_{2}(y_t) \,, \label{eq:WCCHK2}
\end{align}
where the subscript $H^\pm$ denotes the charged-Higgs contributions, $y_q = m^2_q/m^2_{H^\pm}$, and 
the loop integral functions are defined as:
 \begin{subequations}
  \begin{align}
  I^{WH}_1 (y_t, y_W) & =  %%\int^1_0 dx_1 \int^{x_1}_0 dx_2 \frac{x_2}{(1-x_1 +y_t x_2 + y_W (x_1 -x_2))^2} \,, \\
   \frac{1}{(y_t -y_w)^2} \left[\frac{y_t -y_W}{1-y_t } + \frac{y_W \ln y_W}{1-y_W} + \frac{(y^2_t -y_W) \ln y_t }{(1-y_t)^2}\right] \,, \\
  I^{WH}_2 (y_t, y_W) & = %%\int^1_0 dx_1 \int^{x_1}_0 dx_2 \frac{x_2}{1-x_1 +y_t x_2 + y_W (x_1 -x_2)}\,, \\
  -\frac{y_t}{2 (1-y_t) (y_t -y_W)} - \frac{y^2_W \ln y_W}{2(1-y_W) (y_t -y W)^2}  \nonumber \\
  & - \frac{y_t \left(y_t +(2-y_t)y_W \right) \ln y_t}{2(1-y_t)^2 (y_t -y_W)^2} \,, \\
  I^{HH}_1(y_t) & = %%\int^1_0 dx \frac{x(1-x)}{1-x +y_t x }\,, \\
  \frac{1+y_t}{2(1-y_t)^2} + \frac{y_t \ln y_t}{(1-y_t)^3}\,, \\
   I^{HH}_2(y_t) & = %% \int^1_0 dx \frac{x(1-x)}{(1-x +y_t x)^2}\,.
   - \frac{2}{(1-y_t)^2} - \frac{(1 +y_t) \ln y_t }{(1-y_t)^3} \,.
  \end{align}
  \end{subequations}
We note through box diagrams that the couplings of $sbH(A)$ and $dbH(A)$ can induce $\Delta S=2$; however, because the involved quark in the loop is the bottom-quark, the effects should be much smaller than those from the top-quark loop. Here, we ignore their contributions.  

To obtain $\Delta M_K$ and $\epsilon_K$, we define the hadronic  matrix element of $\bar K$-$K$ mixing to be:
\begin{equation}
 M^{*}_{12} = \langle \bar K^0 | {\cal H}^{\Delta S=2} | K^0 \rangle\,. \label{eq:M12}
 \end{equation}
Accordingly, the $K$-meson mixing parameter and indirect CP violating parameter can be obtained as:
\begin{equation}
\Delta M_K \approx  2 Re M_{12} \,,  \quad
\epsilon_K \approx \frac{e^{i \pi/4}}{\sqrt{2} \Delta M_K} Im M_{12}\,, \label{eq:DMeK}
\end{equation}
where we have ignored the small contribution of $Im A_0/ReA_0$ from $K\to \pi \pi$ in  $\epsilon_K$. Since $\Delta M_K$ is experimentally measured well, we will directly take the $\Delta M_K$ data for the denominator of $\epsilon_K$. It has been found that  the short-distance SM result on $\Delta M_K$ can explain the  data by $\sim 70\%$, and the long-distance effects may contribute another~$20-30\%$ with a large degree of uncertainty~\cite{Buras:2014maa}. In this work, we take $\Delta M^{\rm exp}_K$ as an input to bound the new physics effects; using the constrained parameters, we then study the implications on the other phenomena. 

To estimate the $M_{12}$ defined in Eq.~(\ref{eq:M12}), we need to run the Wilson coefficients from a higher scale to a lower scale using the renormalization group (RG) equation. In addition, we also need  the hadronic matrix elements of $\langle\bar K^0 | Q_i | K^0\rangle$. In order to obtain this information, we adopt the results shown in~\cite{Buras:2001ra}, where the RG and nonperturbative QCD effects have been included.  Accordingly, the $\Delta S=2$ matrix element can be expressed as:
 \begin{align}
 \langle\bar K^0 | {\cal H}^{\Delta S=2} | K^0 \rangle & = \frac{G^2_F V_{\rm CKM} }{48 \pi^2} m^2_W m_K f^2_K  \left\{ P^{VLL}_1 \left[ C^{VLL}_{F1}(\mu_t) + C^{VRR}_{F1}(\mu_t)\right] \right. \nonumber \\
 & + P^{LR}_1 C^{LR}_{F1} (\mu_t) + P^{LR}_2 C^{LR}_{F2} (\mu_t)+ P^{SLL}_{1}\left[ C^{SLL}_{F1}(\mu_t) + C^{SRR}_{F1}(\mu_t)\right] \nonumber \\
 & \left. + P^{SLL}_2 \left[ C^{SLL}_{F2}(\mu_t) + C^{SRR}_{F2}(\mu_t)\right]\right\}\,, \label{eq:KK}
 \end{align}
where the Wilson coefficients $C^\chi_{Fi}$ are taken at the $m_t$ scale with $F=S(H^\pm)$, and the values of $P^\chi_i$ at $\mu=2$ GeV are~\cite{Buras:2001ra}:
 \begin{align}
 P^{VLL}_1 & \approx 0.48\,, \quad P^{LR}_1\approx  -36.1\,, \quad  P^{LR}_2 \approx59.3\,, \nonumber \\
 P^{SLL}_1 & \approx -18.1\,, \quad P^{SLL}_2\approx 32.2 \,. 
 \end{align}
 It can be seen that the values of $P^{\chi}_i$, which are related to the scalar operators, are one to two  orders of magnitude  larger than the value of $P^{VLL}_1$, where the enhancement factor is from the factor $m^2_K/(m_s + m_d)^2$. The similar enhancement factor in the $B$-meson system is  just slightly larger than one.  Although the new physics scale is dictated by  $\mu_{S(H^\pm)}$ ($\mu_{S(H^\pm)} > m_t$), as indicted in~\cite{Buras:2001ra}, the RG running of the Wilson coefficients from $\mu_{S(H^\pm)}$ to $m_t$ is necessary only when $\mu_{S(H^\pm)} > 4 m_t$. To estimate the new physics effects, we will take $\mu_{S(H^\pm)} \lesssim 800$ GeV and ignore the running effect between $\mu_{S(H^\pm)}$ and $m_t$ scale. In Eq.~(\ref{eq:KK}), we have explicitly shown the CKM factor to be $V_{\rm CKM}=1$ for $F=S$ and  $V_{\rm CKM}=(V^*_{ts} V_{td})^2$ for $F=H^\pm$.

 \subsection{  $Re(\epsilon'_K/\epsilon_K)$ from the charged-Higgs induced QCD and electroweak penguins} 
 
 Using isospin decomposition, the decay amplitudes for $K\to \pi \pi$ can be written as~\cite{Cirigliano:2011ny}:
  \begin{align}
  A(K^+ \to \pi^+ \pi^0 ) & = \frac{3}{2} A_2 e^{i \delta_2} \,, \nonumber \\
  A(K^0 \to \pi ^+ \pi^- ) & = A_0 e^{i \delta_0} + \sqrt{\frac{1}{2}} A_2 e^{i\delta_2}\,, \nonumber \\
  A(K^0 \to \pi^0 \pi^0 ) & = A_0 e^{i\delta_0} - \sqrt{2} A_2 e^{i\delta_2}\,
  \end{align}
  where $A_{0(2)}$ denotes the isospin $I=0 (2)$ amplitude; $\delta_{0(2)}$ is the strong phase, and the measurement is $\delta_0 - \delta_2 = (47.5 \pm 0.9)^{\circ}$~\cite{Cirigliano:2011ny}.  In terms of the isospin amplitudes, the direct CP violating parameter in $K$ system can be written as~\cite{Buras:2015yba}:
 \begin{equation}
 Re\left( \frac{\epsilon'_K}{\epsilon_K}\right) = - \frac{ a \omega }{\sqrt{2} |\epsilon_K|} \left[ \frac{Im A_0}{ Re A_0} (1 - \hat\Omega_{\rm eff}) - \frac{1}{a} \frac{Im A_2}{ Re A_2}\right]\,,  \label{eq:epsilon_p}
  \end{equation}
 where $a=1.017$~\cite{Cirigliano:2003gt} and $\hat \Omega_{\rm eff}=(14.8 \pm 8.0)\times 10^{-2}$~\cite{Buras:2015yba} include the isospin breaking corrections and the correction of $\Delta I=5/2$, and 
  \begin{equation}
  \omega =  \frac{Re A_2}{ Re A_0} \approx \frac{1}{22.46}\,.
  \end{equation}
With the normalizations of $A_{0,2}$ used in~\cite{Buras:2015yba}, the experimental values of $Re A_0$ and $ReA_2$ should be taken as:
 \begin{equation}
 (ReA_0)^{\rm exp } = 33.22 (1) \times 10^{-8}\, {\rm GeV}\,, \quad  (ReA_2)^{\rm exp} = 1.479(3) \times 10^{-8}\, {\rm GeV}\,.
 \end{equation}
 Although the uncertainty of the  predicted $Re A_0$ in the SM is somewhat  large, the results of $Re A_2$  obtained by the dual QCD approach~\cite{Buras:2014maa} and the RBC-UKQCD collaboration \cite{Boyle:2012ys,Blum:2011ng,Blum:2012uk}  were  consistent  with the experimental measurement. Thus, we can use the  $(Re A_2)^{\rm exp}$ to limit the  new physics effects. Then, the explanation of the measured $Re (\epsilon'_K/\epsilon_K)$ will rely on the new physics effects that contributes to $Im A_0$ and $Im A_2$.

From Eq.~(\ref{eq:Yu_HA}),   the couplings $q q H(A)$ with Cheng-Sher ansatz indeed are suppressed by $m_{d(u)} \tan\beta/v \sim 10^{-3} (\tan\beta/50)$. If we take  $m_{H(A)}$ to be heavier, the effects will be further suppressed. Thus, in the following analysis, we neglect the neutral scalar boson contributions to the $K\to \pi \pi$ processes.  According to the results in~\cite{Chen:2018hqy}, the couplings $u d(s) H^\pm$  as compared with the SM  are  small; therefore, we also drop the tree-level charged-Higgs contributions to $K\to \pi \pi$. Accordingly, the main contributions to the $\epsilon'_K$ are from the top-quark loop QCD and electroweak penguins. Since the induced operators are similar to the SM, in order to consider the RG running of the Wilson coefficients, we thus write the effective Hamiltonian for $\Delta S=1$ in the form of the SM as~\cite{Buchalla:1995vs}: 
 \begin{align}
 {\cal H}^{\Delta S=1} & =  \frac{G_F}{\sqrt{2}} V^*_{us} V_{ud} \sum^{10}_{i=1}  [z_i (\mu)+  \tau y_i (\mu) ]  Q_i (\mu)\,, 
 \end{align}
where  $\tau= -V^*_{ts} V_{td}/(V^*_{us} V_{ud})$, and $z_i(\mu)$ and $y_i(\mu)$ are the Wilson coefficients at the $\mu$ scale. The effective operators for $(V-A)\otimes (V-A)$ are  given as: 
 \begin{equation}
 Q_1 = (\bar s^\alpha u^\beta)_{V-A} (\bar u^\beta d^\alpha)_{V-A}\,, \quad Q_2 = (\bar s u)_{V-A} (\bar u d)_{V-A}\,,
 \end{equation}
 where $\alpha, \, \beta$ are color indices;  the color indices in $\bar q q'$ are suppressed, and $(\bar q q')_{V\pm A} = \bar q \gamma_\mu (1 \pm \gamma_5) q'$. 
For the QCD penguin operators, they are:
 \begin{align}
 Q_3 &= (\bar s d)_{V-A} \sum _q (\bar q q)_{V-A}\,, \ Q_4 =  (\bar s^\alpha d^\beta)_{V-A} \sum _q (\bar q^\beta q^\alpha)_{V-A}\,, \nonumber \\
  Q_5 &= (\bar s d)_{V-A} \sum _q (\bar q q)_{V+A}\,, \ Q_6 =  (\bar s^\alpha d^\beta)_{V-A} \sum _q (\bar q^\beta q^\alpha)_{V+A}\,,
 \end{align}
 where 
 %%the color indices in $\bar q q'$ are suppressed, and $(\bar q' q'')_{V\pm A} = \bar q' \gamma_\mu (1 \pm \gamma_5) q''$, and
  $q$ in the sum includes $u$, $d$, $s$, $c$, and $b$ quarks. 
 For the electroweak penguins, the effective operators are:
  \begin{align}
  Q_7 & =  \frac{3}{2} (\bar s d)_{V-A} \sum _q  e_q (\bar q q)_{V+A}\,, \ Q_8 =  \frac{3}{2} (\bar s^\alpha d^\beta)_{V-A} \sum _q e_q (\bar q^\beta q^\alpha)_{V+A}\,, \nonumber \\
   Q_9 & =  \frac{3}{2} (\bar s d)_{V-A} \sum _q  e_q (\bar q q)_{V-A}\,, \ Q_{10} =  \frac{3}{2} (\bar s^\alpha d^\beta)_{V-A} \sum _q e_q (\bar q^\beta q^\alpha)_{V-A}\,,
  \end{align}
where $e_q$ is the $q$-quark electric charge. The $H^\pm$-mediated Wilson coefficients can be expressed as~\cite{Buchalla:1995vs}:
 \begin{align}
 y^{H^\pm}_3 (\mu_H) & = - \frac{\alpha_{s} (\mu_H)}{24 \pi}E_H(y_t)  + \frac{\alpha C_{H}(x_t,y_t)}{6\pi \sin^2\theta_W}\,, \quad y^{H^\pm}_4 (\mu_H)= \frac{\alpha_s(\mu_H)}{8 \pi} E_H(y_t)  \, , \nonumber \\
 y^{H^\pm}_5(\mu_H) & = - \frac{\alpha_s (\mu_H)}{24 \pi} E_H(y_t)\,, \quad y^{H^\pm}_6 (\mu_H)= \frac{\alpha_s(\mu_H)}{8 \pi} E_H(y_t)\,, \nonumber \\
 y^{H^\pm}_{7}(\mu_H) & = \frac{\alpha}{6 \pi} \left(4 C_{H}(x_t,y_t) + D_H(y_t) \right)\,, \quad  y^{H^\pm}_9(\mu_H) = y^{H^\pm}_{7}(\mu_H) -\frac{4 \alpha C_H(x_t,y_t)}{6\pi \sin^2\theta_W}  \,,
 \end{align}
 $y_8(\mu_H) = y_{10}(\mu_H) =0$, and the functions $D_H$, $C_H$, and $E_H$ are given by~\cite{Davies:1991jt,Buras:2000qz}:
 \begin{align}
 C_{H}(x,y) & =  \zeta^{u*}_{ts}\zeta^{u}_{td}    \left[ \frac{x y}{8(y-1)} - \frac{x y \ln y}{8(y-1)^2} \right] \,, \nonumber \\
 D_{H}(y) &  =  \zeta^{u*}_{ts}  \zeta^{u}_{td} \frac{y}{3} \left[ \frac{47y^2 - 79 y + 38}{36 (y-1)^3} + \frac{-3 y^3 + 6 y -4}{6(y-1)^4} \ln y\right]\,, \nonumber \\
 E_{H}(y) & =  \zeta^{u*}_{ts}\zeta^{u}_{td}   \left[  \frac{y(7 y^2 -29 y +16)}{36 (y-1)^3} + \frac{y (3 y-2)}{6 (y-1)^4} \ln y \right] \,.
   \end{align}
  
 To calculate $\epsilon'_K$, in addition to the Wilson coefficients, we need the hadronic matrix elements of the effective operators. The $Q_{1,2}$ matrix elements  can be obtained from  $Re A_0$ and $Re A_2$ through the parametrizations~\cite{Buras:2015yba}:
   \begin{align}
   Re A_0 & \approx  \frac{G_F}{\sqrt{2}} V^*_{us} V_{ud} \left( z_+ \langle Q_+ \rangle_0 + z_{-} \langle Q_{-} \rangle_0 \right) \nonumber \\
   & = \frac{G_F}{\sqrt{2}} V^*_{us} V_{ud} (1 + q) z_{-} \langle Q_{-} \rangle_0 \,, \nonumber \\
    Re A_2 & \approx  \frac{G_F}{\sqrt{2}} V^*_{us} V_{ud}  z_+ \langle Q_+ \rangle_2\,, 
   \end{align}
  where we ignore the small imaginary part in $V^*_{us} V_{ud}$;  the Wilson coefficients and the matrix elements of the effective operators are taken at the $\mu=m_c$ scale; the subscripts of the brackets denote the isospin states $I=0$ and $I=2$; $z_{\pm} = z_2 \pm z_1$, $Q_\pm = (Q_2 \pm Q_1)/2$, $z_{1}(m_c)=-0.4092$,  $z_2(m_c)=1.2120$, and $q= z_+ \langle Q_+ \rangle_0/ (z_- \langle Q_- \rangle_0$).   In the isospin limit,  the hadronic matrix elements of $Q_{4,9,10}$ can be related to $Q_{+,-}$~\cite{Buras:1993dy}. Therefore, we show the matrix elements  for isospin $I=0$ as~\cite{Buras:1993dy,Buras:2015yba}:
   \begin{align}
   \langle Q_4 \rangle_0 & = 2 \langle Q_-\rangle_0\,, \ \langle Q_9 \rangle_0 = \frac{3}{2} ( \langle Q_+ \rangle _0 - \langle Q_{-} \rangle_0 ) \,, \nonumber \\
   \langle Q_{10} \rangle_0 & =  \frac{3}{2} \langle Q_+ \rangle _0  + \frac{1}{2} \langle Q_{-} \rangle_0 \,, \nonumber \\
   \langle Q_6 \rangle _0 &=   -4 h \left( \frac{m^2_K}{m_s(m_c) + m_d(m_c)}\right)^2 (f_K - f_\pi) B^{(1/2)}_6\,, \nonumber \\
   \langle Q_8 \rangle_0 & =  2 h \left( \frac{m^2_K}{m_s(m_c) + m_d(m_c)}\right)^2 f_\pi B^{(1/2)}_8 \,,
   \end{align}
 for isospin $I=2$, they are given as:
   \begin{align}
   \langle Q_{9} \rangle_ 2 &= \langle Q_{10} \rangle_ 2 = \frac{3}{2} \langle Q_+ \rangle_2\,, \nonumber \\
    \langle Q_8 \rangle_2 & =  \sqrt{2} h \left( \frac{m^2_K}{m_s(m_c) + m_d(m_c)}\right)^2 f_\pi B^{(3/2)}_8\,, 
   \end{align}
   where $h=\sqrt{3/2}$; the small matrix elements for $Q_{3,5,7}$ are neglected; $B^{(1/2)}_6 = 0.57 \pm 0.19$, $B^{(3/2)}_8 =0.76 \pm 0.05$, and $B^{(1/2)}_8 =1.0 \pm 0.2$~~\cite{Buras:2015qea}, which are extracted from the lattice calculations~\cite{Blum:2015ywa,Bai:2015nea}.

  Using the introduced  hadronic matrix elements and Eq.~(\ref{eq:epsilon_p}), the direct CP violating parameter via the charged-Higgs contributions can be expressed as:
   \begin{subequations}
 \begin{align}
 Re\left(\frac{\epsilon'_K}{\epsilon_K} \right)_{H^\pm} & =  Im \left[\lambda_t \left( a( 1 - \hat\Omega_{\rm eff}) P^{(1/2)}_{H^\pm} - P^{(3/2)} _{H^\pm}\right) \right] \,, \\
 P^{(1/2)}_{H^\pm} & = a^{(1/2)}_{ 0}(H^\pm) + a^{(1/2)}_{ 6}(H^\pm) B^{(1/2)}_6\,, \\
 P^{(3/2)}_{H^\pm} & = a^{(3/2)}_{ 0}(H^\pm) + a^{(3/2)}_{8} (H^\pm) B^{(3/2)}_8 \,, 
 \end{align}  
 \label{eq:epdvie}
   \end{subequations}
where $\lambda_t=V^*_{ts} V_{td}$, $a^{\Delta I}_{i}$  are defined as~\cite{Buras:2015yba}:
 \begin{align}
 a^{(1/2)}_0(H^\pm) & \approx r_1 \frac{4 y^{H^\pm}_4 - b (3 y^{H^\pm}_9 - y^{H^\pm}_{10})}{2(1+q) z_{-}} + r_2\, b\, y^{H^\pm}_8 \frac{\langle Q_8 \rangle_0}{Re A_0}\,, \nonumber \\
 a^{(1/2)}_6 (H^\pm) & \approx r_2\, y^{H^\pm}_6 \frac{\langle Q_6 \rangle _0}{B^{(1/2)}_6 Re A_0}\,, \nonumber \\
 a^{(3/2)}_0(H^\pm)  & \approx r_1 \frac{3 (y^{H^\pm}_9 + y^{H^\pm}_{10})}{2 z_+}\,, \nonumber \\
  a^{(3/2)}_8 (H^\pm) & \approx r_2\, y^{H^\pm}_8 \frac{\langle Q_8 \rangle_2}{B^{(3/2)}_8 Re A_2 }\,, \label{eq:a}
 \end{align}    
with $b = 1/(a(1-\hat\Omega_{\rm eff}))$, 
 \begin{align}
r_1 = \frac{\omega}{\sqrt{2} |\epsilon_K| V^*_{us} V_{ud} } \approx 64.545\,, \ r_2 = \frac{\omega G_F}{2  |\epsilon_K|} \approx 1.165 \times 10^{-4}\, \text{GeV}^{-2}\,.
 \end{align}
 We note that the Wilson coefficients in Eq.~(\ref{eq:a}) should be taken at the $\mu=m_c$ scale through the RG running. 

\section{Charged-Higgs on the $K\to \pi \nu \bar \nu$ decays}

 To investigate the new physics contributions to the rare $K$ decays, we adopt the parametrizations shown in~\cite{Buras:2015yca} as:
 \begin{align}
 BR(K^+ \to  \pi ^+ \bar \nu \nu) &= \kappa_+ (1+\Delta_{EM}) \left[ \left(\frac{ImX_{\rm eff}}{\lambda^5} \right)^2 + \left( \frac{Re \lambda_c}{\lambda} P_c (X) + \frac{Re X_{\rm eff}}{\lambda^5}\right)^2\right]\,, \\
 BR(K_L \to \pi^0 \bar \nu \nu) &= \kappa_L \left( \frac{Im X_{\rm eff}}{\lambda^5}\right)^2\,,
 \end{align}
where $\lambda$ is the Wolfenstein parameter; $\Delta_{EM} = -0.003$;  $P_c(X)=0.404 \pm 0.024$ denotes the charm-quark contribution~\cite{Isidori:2005xm,Mescia:2007kn,Buras:2015qea}; $X_{\rm eff}=V^*_{ts} V_{td} [X^{\rm SM}_{L}(K)+X_L(K) + X_R(K)]$  combines the new physics contributions and the SM result of  $X^{\rm SM}_{L}(K)=1.481\pm 0.009$~\cite{Buras:2015yca}, and  the values of $\kappa_{+,L}$ are given as:
 \begin{align}
 \kappa_+ &= (5.173 \pm 0.025)\times 10^{-11} \left( \frac{\lambda}{0.225}\right)^8\,, \nonumber \\
 \kappa_L & =(2.231 \pm 0.013) \times 10^{-10} \left( \frac{\lambda}{0.225}\right)^8\,.
 \end{align}
Here, $X_L(K)$ and $X_R(K)$ denote the contributions from the left-handed and right-handed quark currents, respectively.  

   %%%%
\begin{figure}[phtb]
\includegraphics[scale=0.7]{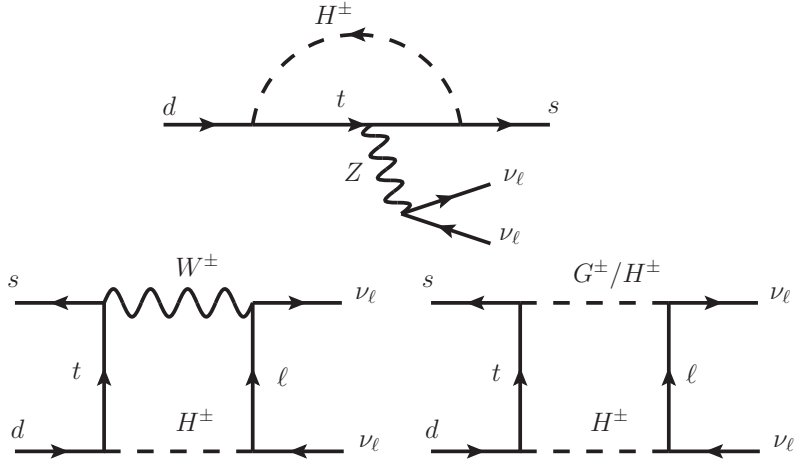}
 \caption{Sketched Feynman diagrams for the $K \to \pi \bar \nu \nu$  process.  }
\label{fig:sdnunu}
\end{figure} 

The neutral scalar bosons $H$ and $A$ do not couple to neutrinos; therefore, the rare  $K\to \pi \nu \bar\nu$ decays can be generated by the Z-mediated electroweak penguins and the $W^\pm$- and $H^\pm$-mediated box diagrams, for which  the representative Feynman diagrams are shown in Fig.~\ref{fig:sdnunu}. Since the dominant $H^\pm$ contributions are from the left-handed quark currents, we only show the $X_L(K)$ results in the following analysis. Using the $H^\pm$ Yukawa couplings in Eq.~(\ref{eq:tqH}), the $Z$-penguin contribution can be obtained as:
\begin{align}
X^{H^\pm}_{L,\rm Pen} & = g^u_L \zeta^{u*}_{ts} \zeta^{u}_{td} x_t y_t J_{1}(y_t)\,, \\
J_1(y_t) & =-\frac{1}{4} \left( \frac{1}{1-y_t} + \frac{\ln y_t}{(1-y_t)^2} \right)\,, \nonumber 
\end{align}
where $g^u_L$ is the $Z$-boson coupling to the left-handed up-type quarks and is given as $g^u_L=1/2 - 2\sin^2\theta_W/3$ with $\sin^2\theta_W\approx 0.23$, and  $J_1(y_t)$ is the loop integral function. 

According to the intermediated states in the loops, there are three types of box diagrams contributing to the $d \to s \nu \bar \nu$ process: $W^\pm H^\pm$, $G^\pm H^\pm$, and $H^\pm H^\pm$. Their results are respectively shown as follows. For the $W^\pm H^\pm$ diagrams, the result is obtained as:
 \begin{align}
 X^{W^\pm H^\pm,\ell}_{L,\rm Box} & = -\frac{1}{\sqrt{3}} (\zeta^{u}_{td} \zeta^{\ell}_{\ell} + \zeta^{u*}_{ts} \zeta^{\ell*}_{\ell})  x_\ell y_{t} t_\beta J_{2}(y_t, y_W)\,, \\
 J_{2}(y_t, y_W) & = I_{2}(y_t, y_W) - I_2(y_t, 0)\,, \nonumber \\
 I_2(y_t, y_W) &= -\frac{1}{4 (y_t -y_W)}  \left[ \frac{2 - y_t }{1-y_t} \ln y_t- \frac{y_W}{1- y_W} \ln y_W\right]\,.
 \end{align}
 where $x_\ell = m^2_\ell/m^2_W$, and the function $J_2$ is  the loop integration. The result of $G^\pm H^\pm$ diagrams is given as:
  \begin{align}
  X^{G^\pm H^\pm, \ell}_{L,\rm Box} & = \frac{1}{\sqrt{3}}( \zeta^{u}_{td} \zeta^{\ell}_{\ell} + \zeta^{u*}_{ts} \zeta^{\ell*}_{\ell})  x_\ell y_{t} t_\beta J_{3}(y_t, y_W)\,, \label{eq:GH}\\
  J_{3}(y_t, y_W) & =\frac{1}{8} \int^1_0 dx_1 \int^{x_1}_0 dx_2 \int^{x_2}_0 dx_3 \frac{1}{1 - (1-y_t) x_1 -(y_t -y_W) x_2 - y_W x_3}\,. \nonumber 
  \end{align}
  It can be seen that except the loop functions,  $X^{G^\pm H^\pm}_{L,\rm Box}$ and $X^{W^\pm H^\pm}_{L,\rm Box}$ have the common factor from the charged-Higgs effect.  The pure $H^\pm$-loop contribution to the box diagram can be written as:
  \begin{align}
  X^{H^\pm H^\pm, \ell}_{L,\rm Box} & = - \frac{1}{\sqrt{3}}\zeta^{u*}_{ts} \zeta^{u}_{td} \left| \zeta^{\ell}_{\ell} \right|^2 x_{\ell} y_t \tan^2\beta J_{4}(y_t)\,, \nonumber \\
  J_{4}(y_t) & = \frac{1}{48} \left[ \frac{1}{1- y_t} + \frac{ y_t \ln y_t }{(1 -y_t)^2}\right] \,. 
  \end{align}
Because the lepton Yukawa coupling is proportional to the lepton mass, all of the box diagrams depend on $x_\ell = m^2_\ell/m^2_W$; therefore, they are dominated by the $\tau$-lepton.  Although the $W^\pm H^\pm$ and $G^\pm H^\pm$ diagrams have a $\tan\beta$ enhancement, the enhancement factor of $H^\pm H^\pm$ is $\tan^2\beta$; that is,  the $H^\pm H^\pm$ contribution overwhelms the  $W^\pm H^\pm$ and $G^\pm H^\pm$.

  \section{ Numerical analysis}
  
  \subsection{ Numerical inputs }% {$\Delta M_K$, $\epsilon_K$ and  $\epsilon'/\epsilon$ in the SM}
  
  The new free parameters considered in this study are $\chi^u_{tt,ct}$, $\chi^d_{bb, bs,bd}$, $\chi^\ell_\tau$, $t_\beta$, and $m_{H,A,H^\pm}$.  In addition to $\Delta M_K$ and $\epsilon_K$,  the charged-Higgs related  parameters also contribute to  the $\Delta M_{B_{d(s)}}$ and $B\to X_s \gamma$ processes, so we have taken these observables into account to constrain the parameters. Thus, the experimental data used to bound the free parameters are~\cite{PDG}:
   \begin{align}
  \Delta M^{\rm exp}_K & \approx 3.48 \times 10^{-15} \text{ GeV} \,,  \ 
     \Delta M^{\rm exp}_{B_d}  =(3.332 \pm 0.0125)\times 10^{-13}\, \text{GeV} \nonumber \\
    \Delta {M^{\rm exp}_{B_s}} & =(1.168 \pm 0.014) \times 10^{-11}\, \text{GeV}\,, \  BR(B\to X_s \gamma)^{\rm exp}   = (3.49 \pm 0.19)\times 10^{-4}\,,  \nonumber \\
    \ \epsilon^{\rm exp}_K  & \approx 2.228\times 10^{-3} \,. 
  \end{align}
Since $\epsilon_K$ in the SM fits well with the experimental data~\cite{Buchalla:1995vs}, we use
 \begin{equation}
 \epsilon^{\rm NP}_K = \kappa_\epsilon \times 10^{-3} \ \text{with} \ |\kappa_\epsilon| < 0.4  \label{eq:eNP_K}
 \end{equation} 
 to constrain the new physics effects~\cite{Buras:2015jaq}.  The uncertainties of NLO~\cite{Herrlich:1993yv} and NNLO~\cite{Brod:2011ty} QCD corrections to the short-distance contribution to $\Delta M_K$ in the SM are somewhat large, we take the combination of  the short-distance (SD) and long-distance (LD) effects as $\Delta M^{\rm SM}_K (SD + LD)    = (0.80 \pm 0.10) \Delta M^{\rm exp}_K$~\cite{Buras:2014maa}. Accordingly, the new physics contribution to $\Delta M_K$ is limited to: 
  \begin{equation}
   \Delta M^{\rm NP}_K= r_K \Delta M^{\rm exp}_K \ \text{with} \ |r_K|< 0.2 \,. \label{eq:DMNP_K}
  \end{equation}

Although the charged-Higgs can generally contribute to  the tree-level processes and contaminate the determination of CKM matrix elements, the charged-Higgs-induced tree processes indeed can be ignored because these processes, which can determine  $V_{ud,cs}$, $V_{ud, cd}$, and $V_{ub, cb}$, will be suppressed by $1-\chi^\ell_\ell/s_\beta$ when $\chi^\ell_\ell\sim 1$ are taken. Then, the CKM matrix elements mentioned above can be taken  the same as those obtained in the SM. With the Wolfenstein parametrization~\cite{Wolfenstein:1983yz},  the CKM matrix elements can be taken as:
 \begin{align}
 V_{ud} & \approx V_{cs} \approx 1-\lambda^2/2\,, \ V_{us}\approx - V_{cd}\approx \lambda = 0.225\,, \nonumber\\
 V_{cb} &\approx  0.0407\,, V_{ub}\approx 0.0038 e^{-i\phi_3}\,, \ \phi_3=73.5^{\circ}\,, \label{eq:CKM}
 \end{align}
where $V_{cb}$ and $V_{ub}$ are taken from the averages of inclusive and exclusive semileptonic decays~\cite{Buras:2015qea}, and the $\phi_3$ angle is the central value averaged by the heavy flavor averaging group (HFLAV) through all  charmful two-body $B$-meson decays~\cite{Amhis:2016xyh}. In terms of the CKM matrix elements shown in Eq.~(\ref{eq:CKM}), we can obtain the other CKM matrix elements and CP phase of $V_{td}$ as~\cite{Buras:2015qea}:
 \begin{align}
R_b &\approx \left(1- \frac{\lambda^2}{2}\right) \frac{|V_{ub}|}{\lambda |V_{cb}|} \approx 0.40\,, \nonumber \\
 R_{t}& =\left( 1+ R^2_b - 2R_b \cos\phi_3 \right)^{1/2} \approx 0.96\,, \ V_{ts}\approx -V_{cb}\,, \ V_{tb}\approx 1\,, \nonumber \\
 |V_{td}|& = |V_{us}| |V_{cb}| R_t  \approx 0.0088\,, \ \phi_2={\rm arccot} \left(\frac{1-R_b\cos\phi_3}{R_b \sin\phi_3}\right)\approx 23.4^{\circ}\,,
 \end{align}
%The values of the CKM matrix elements are taken as follows:
 %\begin{align}
% V_{ud} & \approx V_{cs} \approx 1-\lambda^2/2\,, \ V_{us}\approx - V_{cd}\approx \lambda = 0.225\,, \nonumber \\
% V_{td} & \approx 0.0088 e^{-i \phi_2}\,, \ \phi_2 \approx 23^{\circ}\,, \ V_{ts} \approx -0.041 \,, \ V_{tb}\approx 1\,, 
% \end{align}
 where $Re(V^*_{ts} V_{td}) \approx  -3.3\times 10^{-4}$ and $Im(V^*_{ts} V_{td}) \approx 1.4 \times 10^{-4}$ are close to  those values  used in~\cite{Buras:2015yba}.

 The particle masses used to estimate the numerical values are given as:
  \begin{align}
  & m_K\approx 0.489 \text{ GeV}\,,~m_{B_d}\approx 5.28 \text{ GeV}\,,~m_{B_s}\approx 5.37 \text{ GeV}\,,~m_W \approx 80.385 \text{ GeV} \,, \nonumber \\
  & m_t \approx 165 \text{ GeV}\,, ~m_c\approx 1.3 \text{ GeV}\,, ~m_s(m_c) \approx  0.109 \text{ GeV}\,, ~m_d (m_c) \approx 5.44 \text{ MeV}\,.
  \end{align}
In the 2HDM, the mass spectra of $H$, $A$ and $H^\pm$ are not  independent parameters, and they are correlated through the parameters in the Higgs potential and constrained by the vacuum stability, Peskin-Takeuchi parameters~\cite{Peskin:1991sw}, and Higgs precision measurements~\cite{Benbrik:2015evd}. Following the results in~\cite{Benbrik:2015evd}, the maximum mass difference  $|m_{H(A)}-m_{H^\pm}|$ should be around 100 GeV when $m_{H}=m_{A}$ is adopted. Since the interesting region of $m_{H^\pm}$ in this study is near $200$ GeV, we take $m_{H(A)}=300$ GeV  GeV as the input to show the numerical analysis.

\subsection{ Direct bound on $m_{H^\pm}$ from the LHC}
  
  According to Eq.~(\ref{eq:Clepton}), if we take $\chi^\ell_\tau=1$, the $H^\pm$ Yukawa coupling to the tau-lepton is suppressed by $1-\chi^\ell_\tau/s_\beta$; therefore, in our case, the charged-Higgs with $m_{H^\pm} \sim 200$ GeV predominantly decays to the $t \bar b$ final state.  Thus, the experimental limit on $m_{H^\pm}$ is from the CMS data at $\sqrt{s}=8$ TeV~\cite{Khachatryan:2015qxa}, and the upper bound  for $m_{H^\pm}=200$ GeV and $BR(H^+\to t \bar b)=1$ is $\sigma(pp\to \bar t (b) H^\pm) < 1.53$ pb~\cite{Krawczyk:2017sug}, where the cross section  $\sigma(pp\to \bar t (b) H^\pm)$  includes the $pp\to\bar t(b) H^+$ and $pp\to  t(\bar b) H^-$ contributions. From the Yukawa  sector, the $ t b H^\pm$ couplings can be expressed as:
  \begin{align} 
  {\cal L}^{H^\pm}_{Y} & \supset    \frac{\sqrt{2}}{v} V_{tb}  \bar t \left( m_t   \zeta^u_{tb}  P_L  + m_b \zeta^d_{tb} P_R \right)b + H.c. \,, \nonumber \\
\zeta^u_{tb} &=\frac{1}{t_\beta}  - \frac{\chi^{u*}_{tt}}{s_\beta}\,, \  \zeta^d_{tb} = t_\beta \left( 1 - \frac{\chi^d_{bb}}{s_\beta} \right)\,.
 \end{align}
Accordingly, the main charged-Higgs production channel is $gg \to \bar t b H^+$ in four-flavor scheme (4FS) and is $g \bar b \to  \bar t H^+$ in five-flavor scheme (5FS), where the 4FS and 5FS are used  to avoid double counting, which happens when the $b$-quark final state in $gg \to \bar t b H^+$ escapes detection~\cite{DiazCruz:1992gg,Borzumati:1999th,Akeroyd:2016ymd}.   

\begin{figure}[phtb]
\includegraphics[scale=0.5]{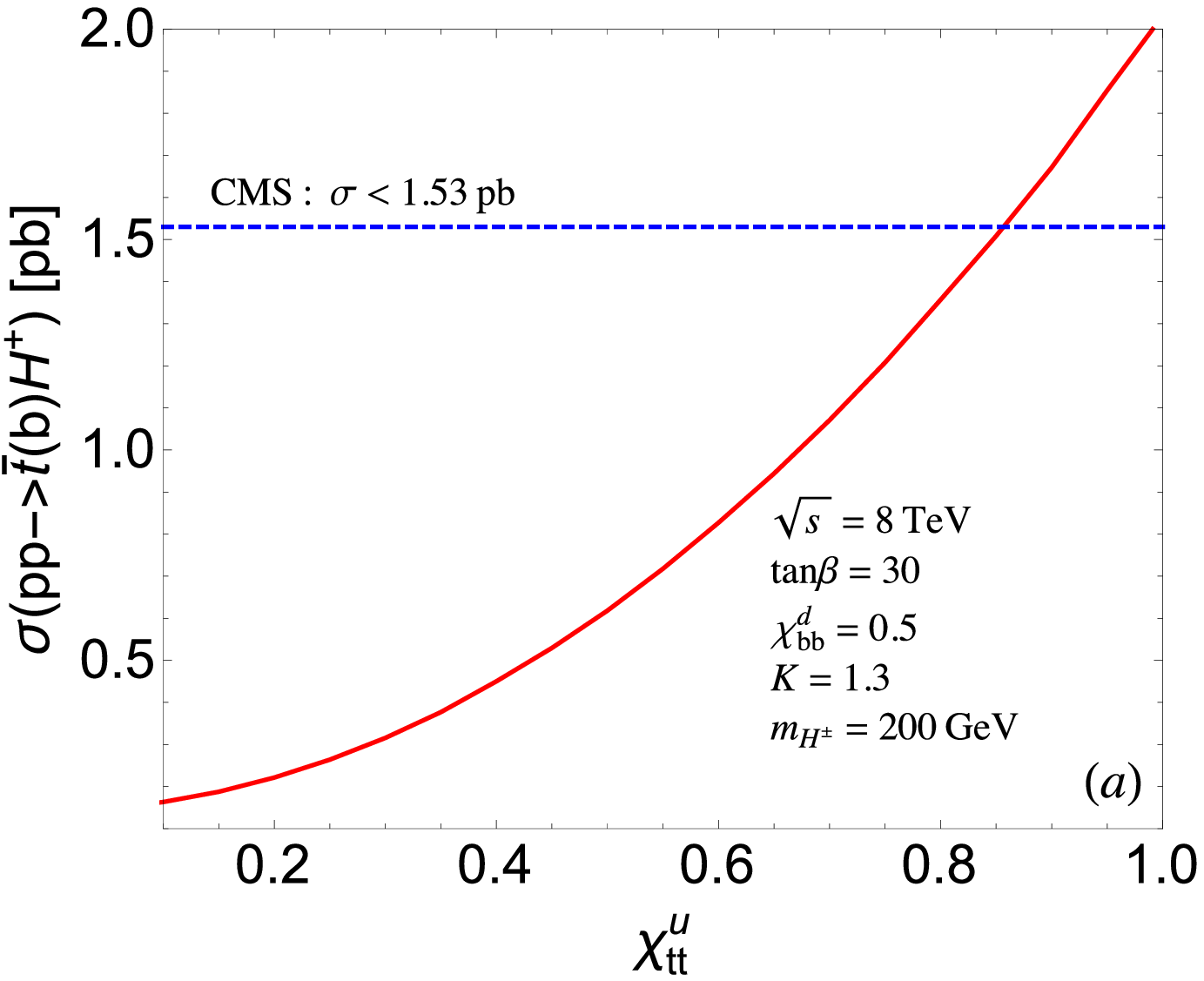}
\includegraphics[scale=0.5]{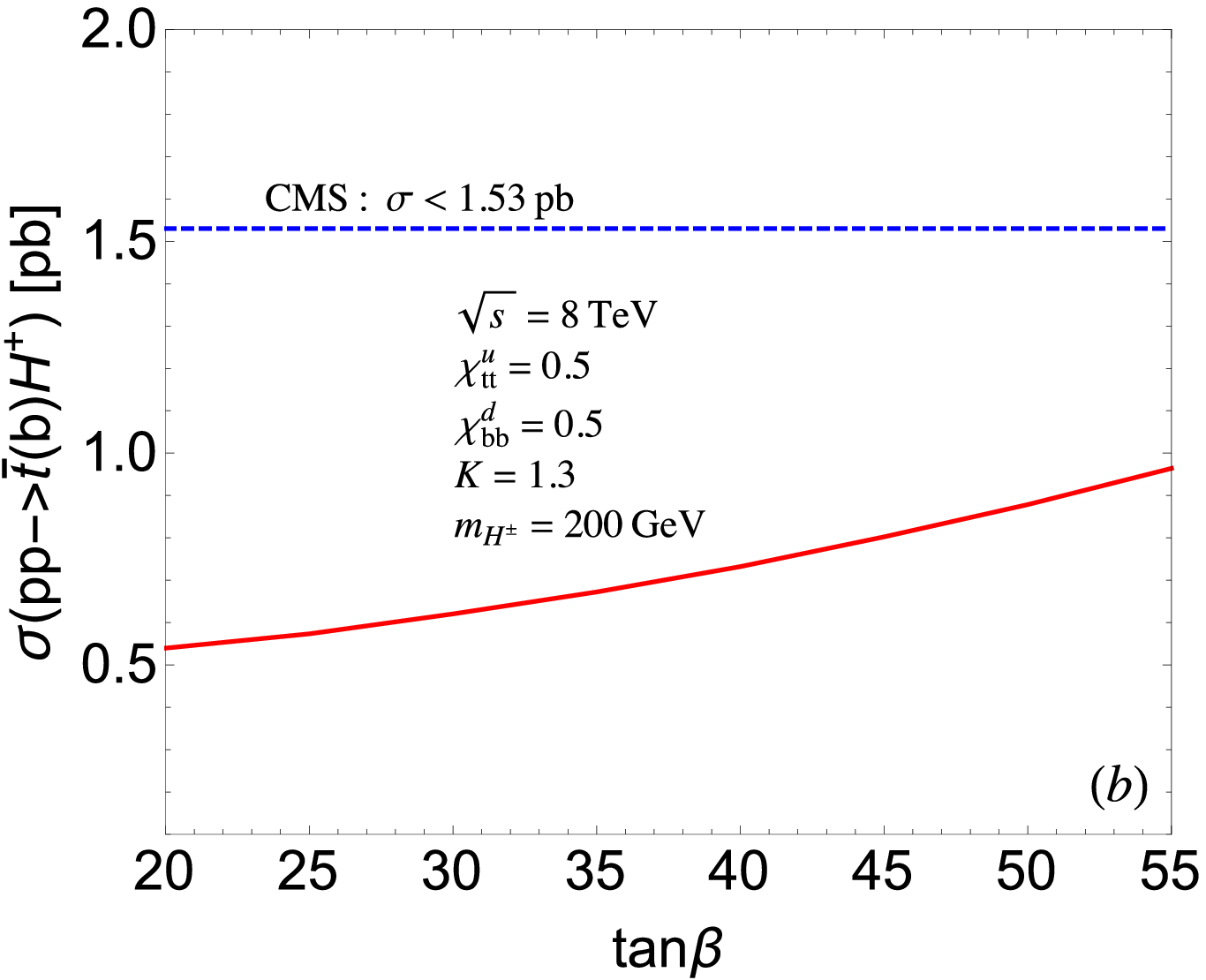}

 \caption{ Production cross-section of $pp\to \bar t(b) H^+$ (solid)  at $\sqrt{s}=8$ TeV as a function of (a) $\chi^u_{tt}$ and (b) $t_\beta$, where  the  fixed values of parameters are given in the plots, $K=1.3$ is  the $K$-factor for the  radiative QCD corrections, and the dashed line denotes the CMS upper bound.}
\label{fig:mH_limit}
\end{figure}

To estimate the production cross-section for $pp\to \bar t (b) H^\pm$, we employ  CalcHEP~\cite{Belyaev:2012qa} associated with  the {\tt CTEQ6L} parton distribution functions (PDFs)~\cite{Nadolsky:2008zw}.
Using $t_\beta=30$, $\chi^d_{bb}=0.5$, and $m_{H^\pm}=200$ GeV, the $\sigma(pp\to \bar t (b) H^\pm)$  at $\sqrt{s}=8$ TeV as a function of $\chi^u_{tt}$ is shown in Fig.~\ref{fig:mH_limit}(a), where the dashed line is the CMS upper limit, and $K=1.3$ denotes the K-factor for the radiative  QCD corrections~\cite{Plehn:2002vy}. The dependence of $\sigma(pp\to \bar t (b) H^\pm)$ on $t_\beta$ is shown in Fig.~\ref{fig:mH_limit}(b), where $\chi^u_{tt}=0.5$, $\chi^d_{bb}=0.5$, $m_{H^\pm}=200$ GeV, and $K=1.3$ are used. From the plots, it can be seen that taking proper values of $\chi^u_{tt}$ and $\chi^d_{bb}$, both $m_{H^\pm}\sim 200$ GeV and  large $t_\beta$ value  can still satisfy the upper limit from the direct search. 

  \subsection{Constraints from $\Delta M_{K,B}$, $B\to X_s \gamma$,  and $\epsilon_K$ }
  
\begin{figure}[phtb]
\includegraphics[scale=0.6]{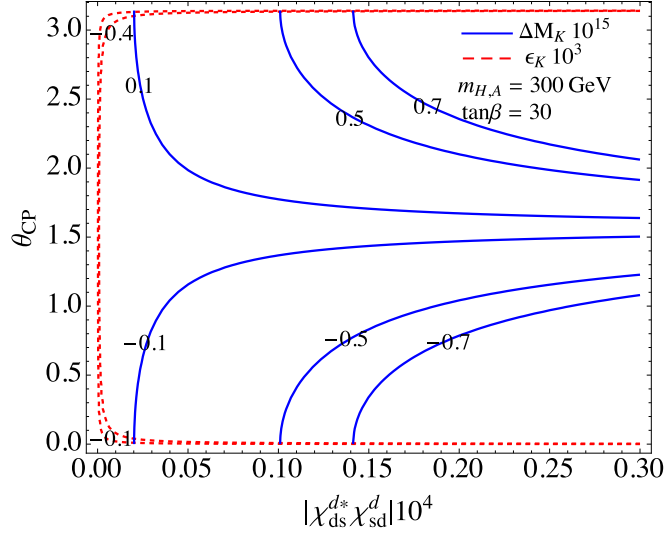}
 \caption{Contours for the $H/A$-mediated  $\Delta M_K$ (in units of $10^{-15})$ and $\epsilon_K$ (in units of $10^{-3}$) as a function of $|\chi^{d*}_{ds}\chi^d_{sd}|$ scaled by $10^{-4}$ and $\theta_{CP}$. }
\label{fig:DMKeK_S}
\end{figure}

   Since the free parameters for the tree-induced  $\Delta S=2$ are  different from those that are  box-induced, we analyze them separately. According to Eq.~(\ref{eq:WCSK2}), in addition to the $t_\beta$ parameter, the main parameters in the $H/A$-mediated $M_{12}$ are $\chi^{d*}_{ds}\chi^d_{sd}=|\chi^{d*}_{ds}\chi^d_{sd}| e^{-i\theta_{CP}}$, where $\theta_{CP}$ is the weak CP-violation phase.  Using Eq.~(\ref{eq:KK}) and  the taken input values, $\Delta M_K$ (solid) and $\epsilon_K$ (dashed)  as a function of $|\chi^{d*}_{ds}\chi^d_{sd}|$ (in units of $10^{-4}$) and $\theta_{CP}$ are shown in Fig.~\ref{fig:DMKeK_S}, where we only show the range of $\theta_{CP}=[0,\, \pi]$ and fix $t_\beta=30$.  It is seen that the typical value of $|\chi^d_{ds, sd}|$ constrained by the $K^0-\bar K^0$ mixing  is $\sim 4.5\times 10^{-3}$. Because the $\epsilon_K$ and $\Delta M_K$ both arise from the same complex parameter $\chi^{d*}_{ds} \chi^{d}_{sd}$, to obtain  $\epsilon_K$ of $O(10^{-3})$, the CP-violation phase $\theta_{CP}$ inevitably has to be of $O(10^{-3})$ away from zero or $\pi$ when the $|\chi^{d*}_{ds} \chi^{d}_{sd}|$ of $O(10^{-5})$ is taken. Intriguingly, the small $\theta_{CP}$ may not be a fine-tuning result in the case of $m_{H}=m_{A}$. If   the Yukawa matrices $Y^f_i$ in Eq.~(\ref{eq:hHHA})   are  symmetric matrices, due to $V^f_R=V^{f*}_L$,  ${\bf X}^f$ in Eq.~(\ref{eq:Xfs})  also being symmetric,  
   we can obtain $\chi^{d}_{ds} = \chi^d_{sd}$ and $\theta_{CP}=0$. Hence, a small $\theta_{CP}$ can be ascribed to a slight break  in a symmetric Yukawa matrix. 
     %%%%

Next, we analyze the charged-Higgs loop contributions to $\Delta M_K$ and $\epsilon_K$. According to Eqs.~(\ref{eq:tqH}) and  (\ref{eq:zetas}), the relevant Yukawa couplings are $\chi^u_{tt,ct}$ and  $\chi^d_{bd, bs}$. However, the same parameters  also contribute to $\Delta M_{B_{d,s}}$ and $B\to X_s \gamma$, where the former can arise from the tree $H/A$-mediated and  charged-Higgs-mediated box diagrams, and the latter is from the $H^\pm$-penguin loop diagrams~\cite{Chen:2018hqy}. Thus, we have to constrain the free parameters by taking the $\Delta M_{B_d, B_s}$ and $B\to X_s \gamma$ data into account. To scan the parameters, we set  the ranges of the scanned  parameters to be:
\begin{align}
& -1 \leq  \chi^{u}_{tt},\, \chi^u_{ct},\, \chi^d_{bb} \leq 1\,, ~
  -0.05\leq  \chi^d_{bd}\,, \chi^{d}_{bs} \leq 0.05\,, \nonumber \\
 & 20 \leq t_\beta \leq 50\,, ~ 200 \leq m_{H^\pm}\; [\text{GeV}] \leq 230\,.
 \end{align}
$\epsilon'_K/\epsilon_K$ can be significantly enhanced  only by the light $H^\pm$; therefore, we set $m_{H^\pm}$ around 200 GeV.

\begin{figure}[phtb]
\includegraphics[scale=0.65]{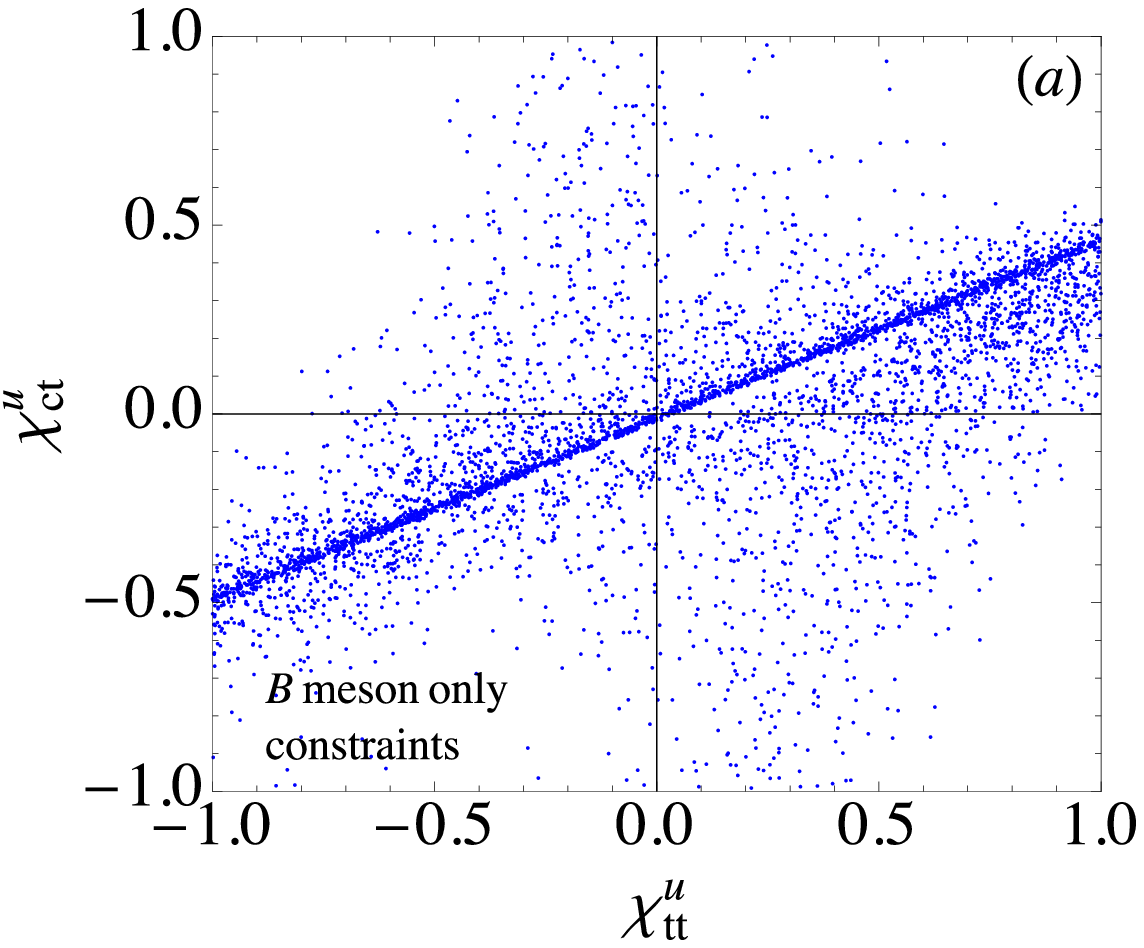}
\includegraphics[scale=0.65]{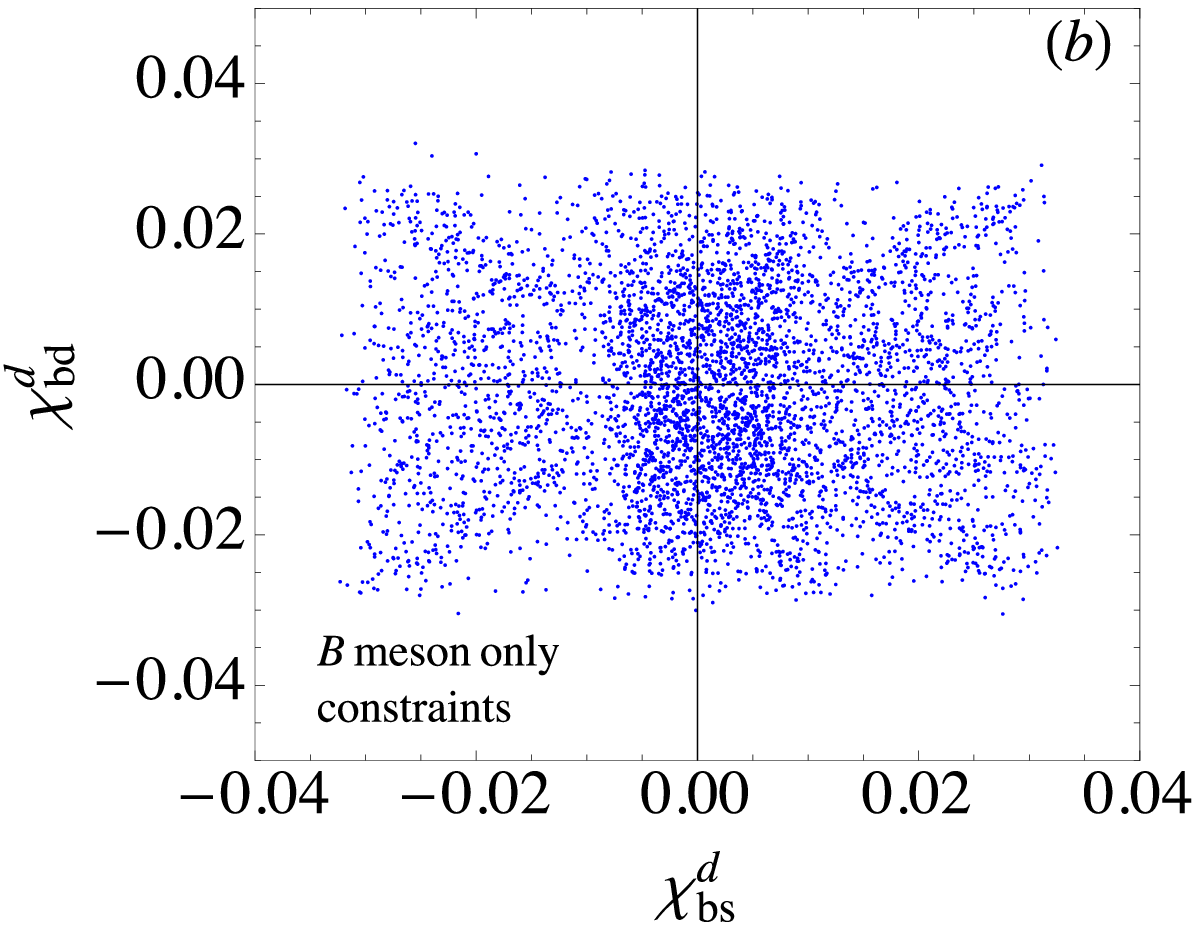}
 \caption{ Constraints from $\Delta M_{B_d, B_s}$ and $B\to X_s \gamma$, where the plots (a) and (b) denote the allowed ranges of $\chi^u_{tt}$ and $\chi^u_{ct}$ and the allowed ranges of $\chi^d_{bs}$ and $\chi^d_{bd}$, respectively. The number of sample points used for the scan is $5\cdot 10^{6}$. }
\label{fig:B_limit}
\end{figure}

\begin{figure}[phtb]
\includegraphics[scale=0.6]{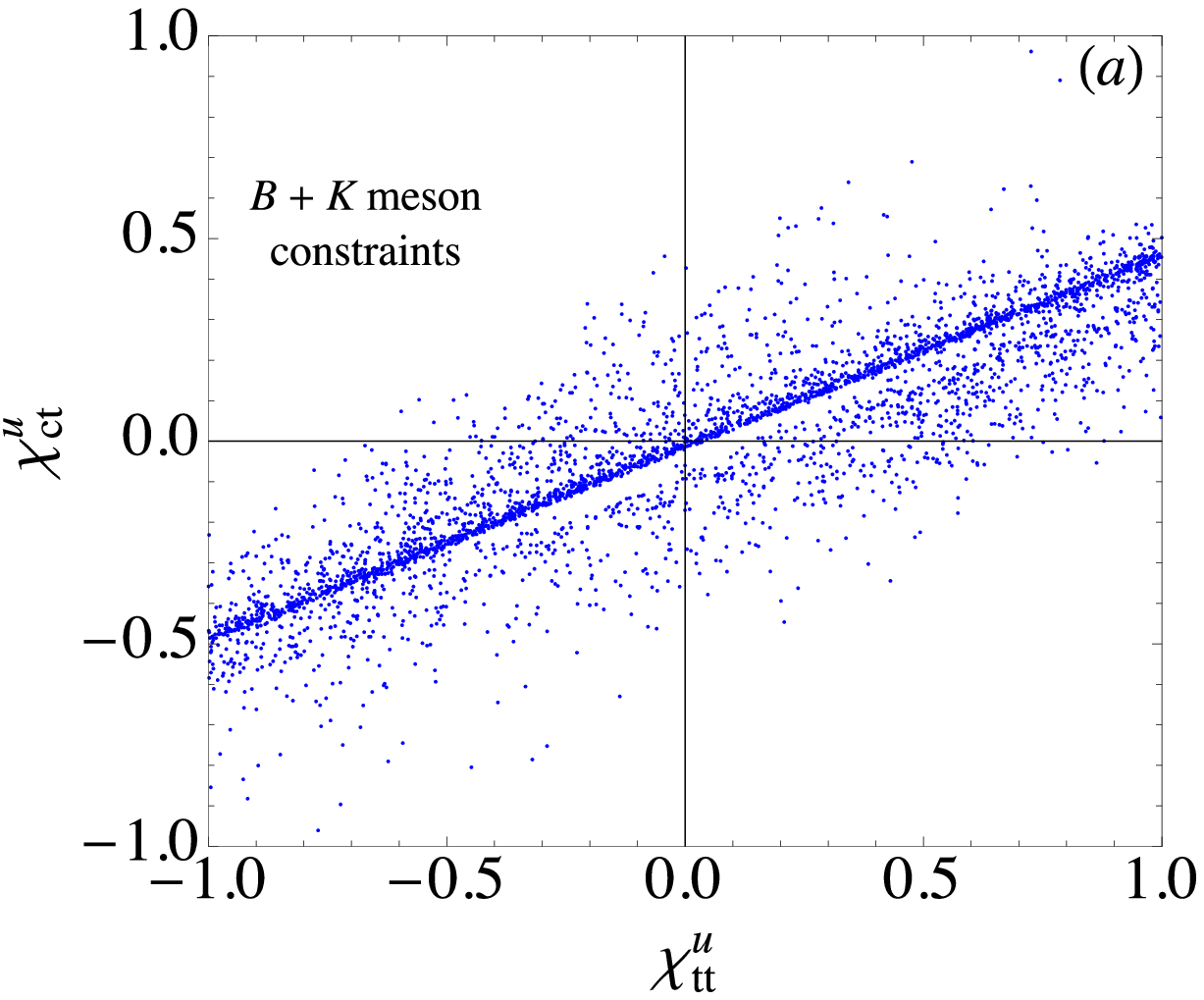}
\includegraphics[scale=0.6]{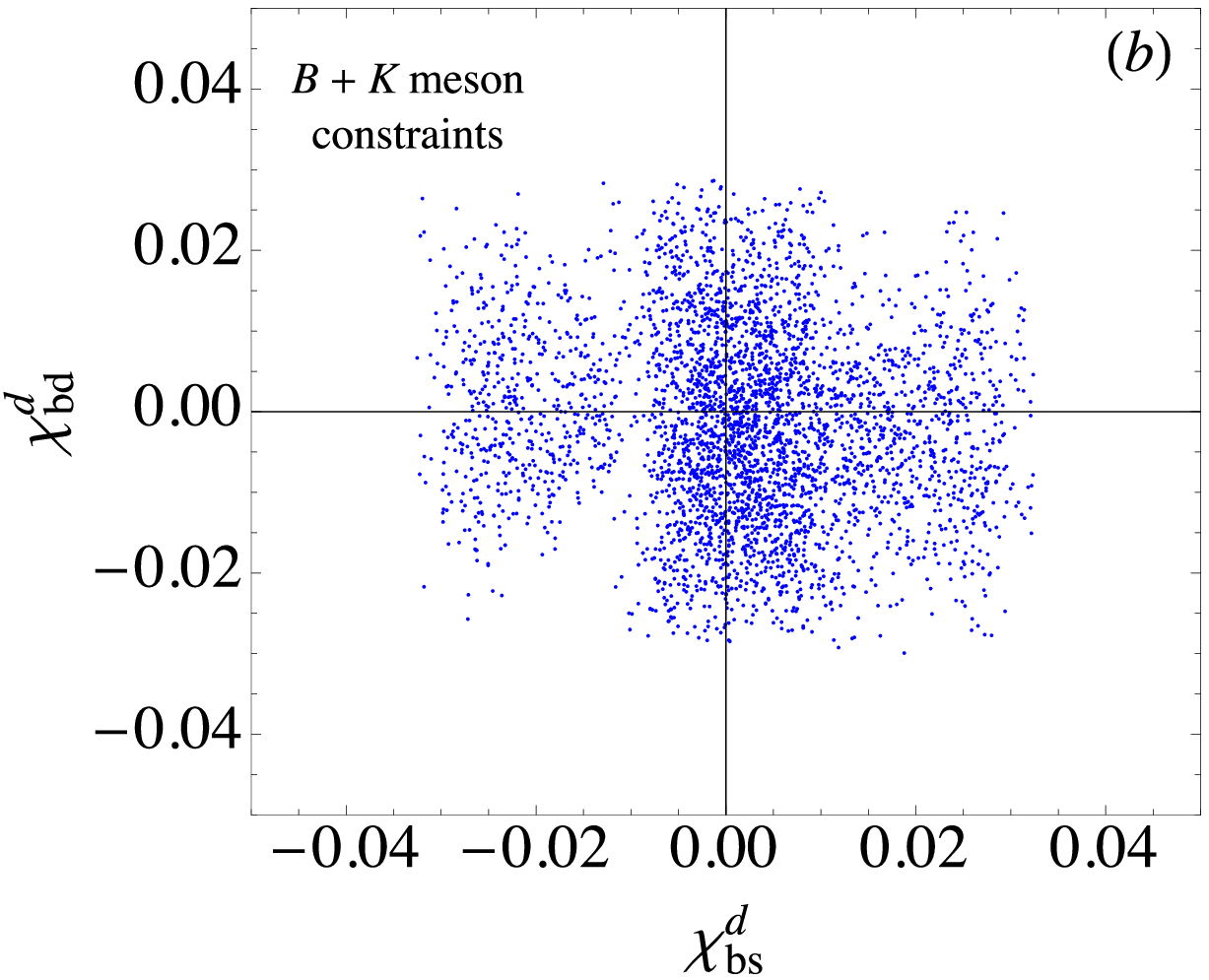}
 \caption{ Based on the results in Fig.~\ref{fig:B_limit}, the constraints from the $H^\pm$-induced $\Delta M_K$ and $\epsilon_K$ are included. }
\label{fig:BpK_limit}
\end{figure}

In order to consider the constraints from the $B$-meson decays, we use the formulae and results obtained in~\cite{Chen:2018hqy}. To understand the influence of $B$ and $K$ systems on the parameters, we show the constraints with and without the $\Delta M_K$ and $\epsilon_K$ constraints.  Thus, for the $\Delta M_{B_{d}, B_{s}}$ and $B\to X_s \gamma$ constraints only, we respectively show the allowed ranges of $\chi^u_{tt}$ and $\chi^{u}_{ct}$ and the allowed ranges of $\chi^d_{bs}$ and $\chi^d_{bd}$ in Fig.~\ref{fig:B_limit}(a) and (b), where the sampling data points for the scan are $5\cdot 10^{6}$. When the $\Delta M_K$ and $\epsilon_K$ constraints are included, the corresponding situations are shown in Fig.~\ref{fig:BpK_limit}(a) and (b), respectively. From the plots, it can be clearly seen that $K$-meson data can further constrain the free parameters.  We note that the results of Fig.~\ref{fig:BpK_limit} do not include the tree-induced $\Delta S=2$ because the involved parameters are different.  

It was studied that the charged-Higgs and neutral Higgs bosons $h$, $H$ and $A$ can also have  significant contributions to the rare  $B_s \to \mu^+ \mu^-$ decay through loop and tree Feynman diagrams in the type-III 2HDM~\cite{Logan:2000iv,Huang:2000sm,Chankowski:2000ng,Bobeth:2001sq} when a large $t_\beta$ scheme is applied, where the current LHCb measurement is $BR(B_s \to \mu^+ \mu^-)=(3.0\pm 0.6^{+0.3}_{-0.2})\times 10^{-9}$~\cite{Aaij:2017vad} and  the SM prediction is $BR(B_s \to \mu^+ \mu^-)=(3.65\pm 0.23 )\times 10^{-9}$~\cite{Bobeth:2013uxa}. The small difference in  $BR(B_s \to \mu^+ \mu^-)$ between experimental and theoretical result leads to a strict constraint on the new physics contribution.   From the Yukawa couplings shown in Eqs.~(\ref{eq:HAY}) and (\ref{eq:CHY}), it can be seen that under the alignment limit (i.e., $\beta-\alpha=\pi/2$), which we adopt in the paper,  $H^\pm$, $H$, and $A$ bosons coupling to the leptons are all proportional to $1-\chi^\ell_{\ell}/s_\beta$ when the Cheng-Sher ansatz is applied. That is, the contributions to the $B_s \to \mu^+ \mu^-$ decay, which arise from the $H^\pm$-induced box diagrams and $H(A)$-penguin and $H(A)$-tree diagrams, can be suppressed in the type-III 2HDM when $\chi^\ell_\mu \sim 1$ is taken.  The choice of $\chi^\ell_\mu \sim 1$ matches with the condition  of $\chi^\ell_\ell\sim 1$ used for taking the SM CKM matrix elements as the numerical  inputs. Hence, in this study, we do not include the constraint from the $B_s \to \mu^+ \mu^-$ decay.

  \subsection{$Re(\epsilon'_K/\epsilon_K)$ in the 2HDM}
  In this subsection, we analyze the charged-Higgs effect on $Re(\epsilon'_K/\epsilon_K)_{H^\pm}$ in detail. 
  To estimate  $Re(\epsilon'_K/\epsilon_K)_{H^\pm}$, we need to run the Wilson coefficients from the $\mu_H$ scale to the $\mu=m_c$ scale.  For new physics effects, we use the LO QCD corrections to the QCD and electroweak penguin operators~\cite{Buchalla:1995vs}; as a result, the relevant Wilson coefficients at the $\mu=m_c$ scale can be obtained as:
  \begin{align}
  y^{H^\pm}_4 (m_c)& \approx -0.6 y^{H^\pm}_3 (\mu_H) +1.07 y^{H^\pm}_4(\mu_H) + 0.08 y^{H^\pm}_5(\mu_H) \nonumber \\
  &+0.46 y^{H^\pm}_6(\mu_H)+0.016 y^{H^\pm}_7(\mu_H)+0.068 y^{H^\pm}_9(\mu_H)\,,  \nonumber \\
    y^{H^\pm}_6 (m_c)& \approx -0.1 y^{H^\pm}_3 (\mu_H) +0.388 y^{H^\pm}_4(\mu_H) + 0.794 y^{H^\pm}_5(\mu_H) \nonumber \\
  &+2.872 y^{H^\pm}_6(\mu_H)+0.02 y^{H^\pm}_7(\mu_H)+0.1 y^{H^\pm}_9(\mu_H)\,, \nonumber \\
    y^{H^\pm}_8 (m_c) & \approx 0.904 y^{H^\pm}_7(\mu_H) \,, \    %
    y^{H^\pm}_9 (m_c)  \approx 1.31 y^{H^\pm}_9(\mu_H) \,, \nonumber \\
 y^{H^\pm}_{10} (m_c) & \approx -0.58 y^{H^\pm}_9(\mu_H) \,,
  \end{align}
  where the $\Lambda$ scale in $\alpha_s$ is determined by $\alpha_s^{(f=5)}(m_Z)=0.118$. 
  
\begin{figure}[phtb]
\includegraphics[scale=0.6]{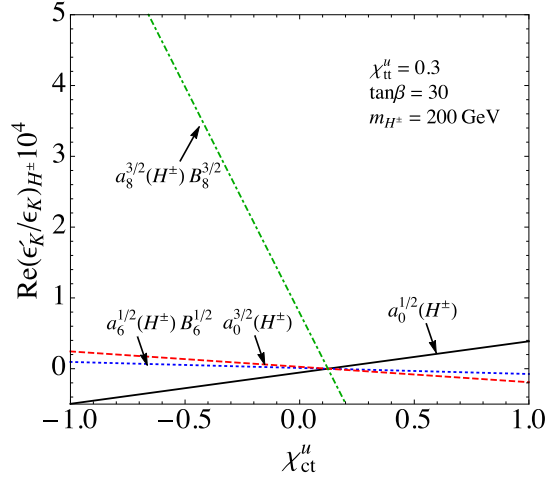}
 \caption{ Each contribution of $a^{1/2}_0(H^\pm)$, $a^{1/2}_6(H^\pm) B^{(1/2)}_6$, $a^{3/2}_0(H^\pm)$, and $a^{3/2}_8(H^\pm)B^{3/2}_8$ defined in Eq.~(\ref{eq:a}), where we fixed $\chi^u_{tt}=0.3$, $t_\beta=30$, and $m_{H^\pm}=200$ GeV.}
\label{fig:epp_P12_P32}
\end{figure}

  According to the parametrization of $Re(\epsilon'_K/\epsilon_K)$ defined in Eq.~(\ref{eq:epdvie}), four pieces contribute to the direct CP violation; two of them, $a^{(1/2)}_0(H^\pm)$ and $a^{(1/2)}_6(H^\pm) B^{(1/2)}_6$, are from $P^{(1/2)}_{H^\pm}$, and the other two, $a^{(3/2)}_0(H^\pm)$ and $a^{(3/2)}_8(H^\pm) B^{(3/2)}_8$, are from $P^{(3/2)}_{H^\pm}$.  To understand their contributions to $\epsilon'_K/\epsilon_K$, we show each individual effect as a function of $\chi^u_{ct}$ in Fig.~\ref{fig:epp_P12_P32}, where for numerical illustration, we have fixed $\chi^{u}_{tt}=0.3$, $t_\beta=30$, and $m_{H^\pm}=200$ GeV. From the results, it can be seen that $a^{3/2}_8 B^{3/2}_8$, which arises from the electroweak penguin $Q_8$ operator, dominates the others.  
  
\begin{figure}[phtb]
\includegraphics[scale=0.5]{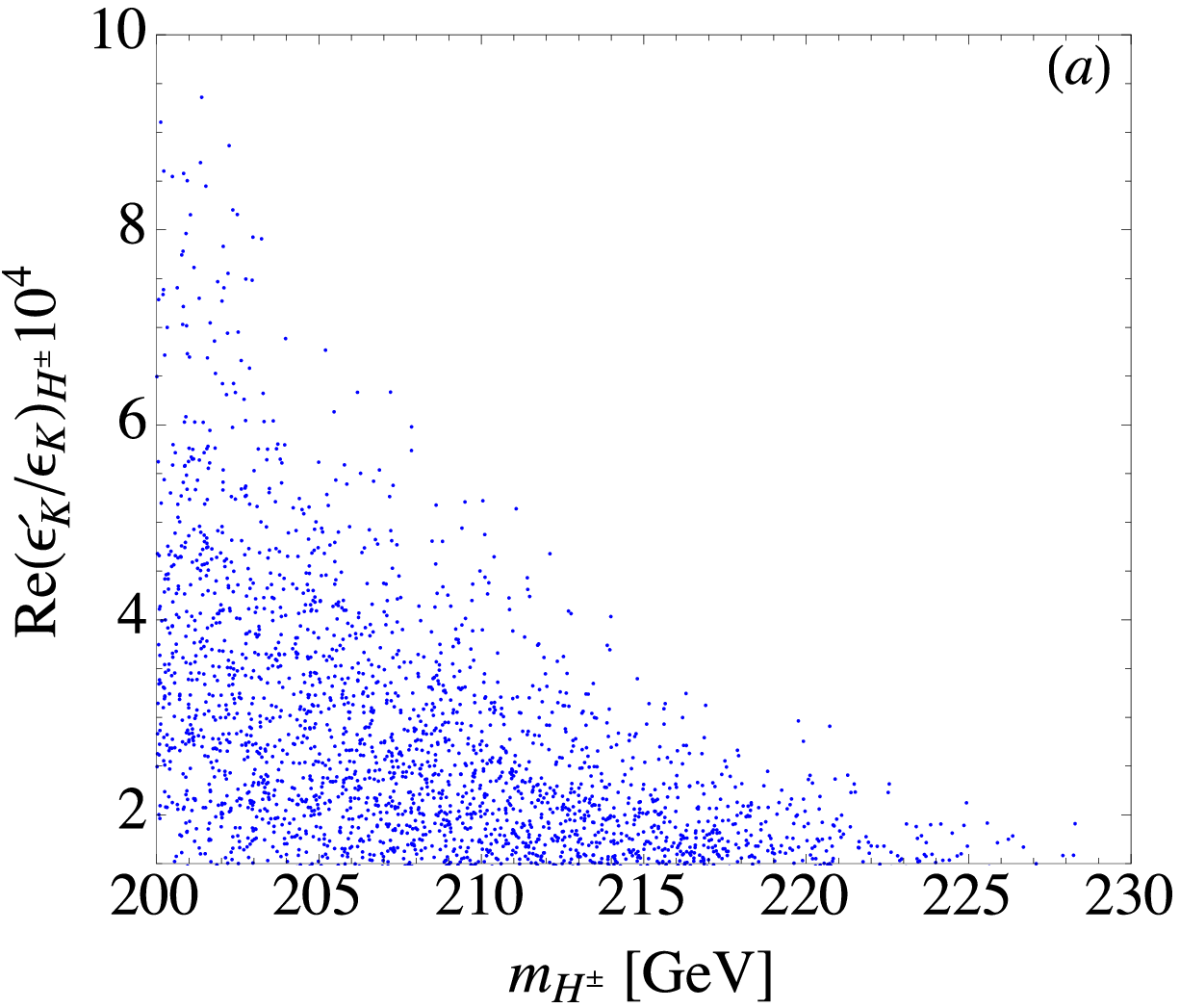}
\includegraphics[scale=0.5]{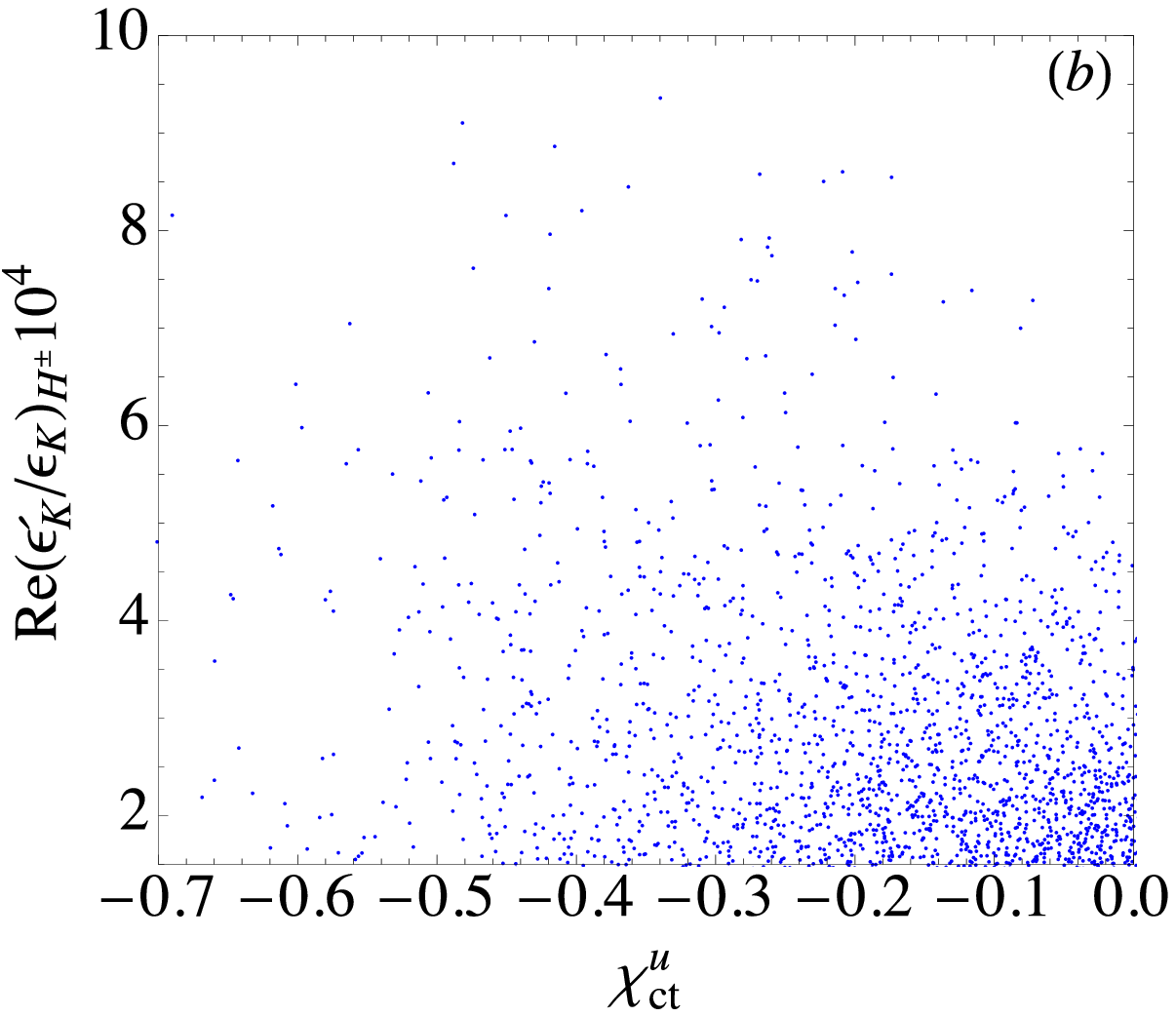}
\includegraphics[scale=0.5]{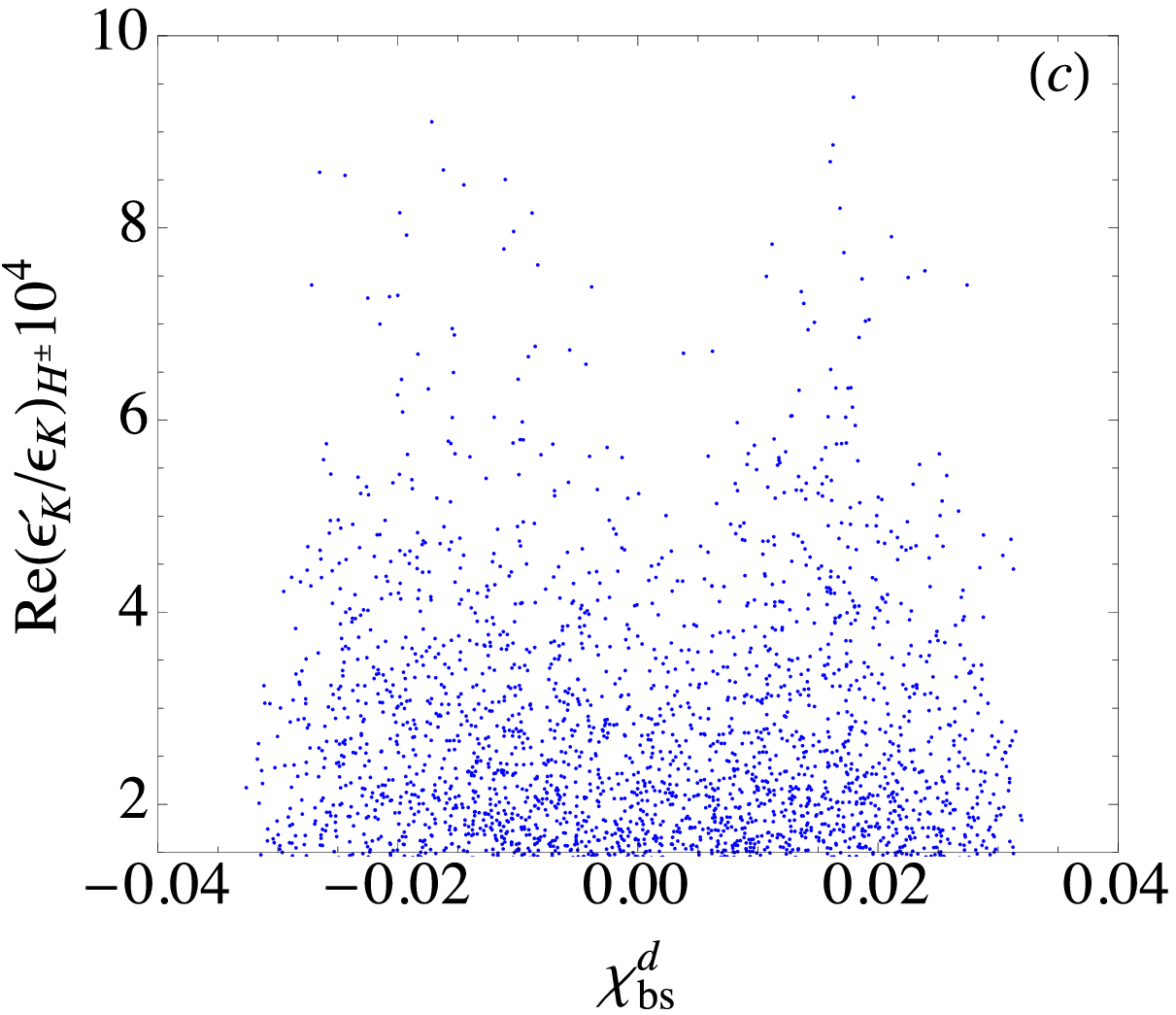}
\includegraphics[scale=0.5]{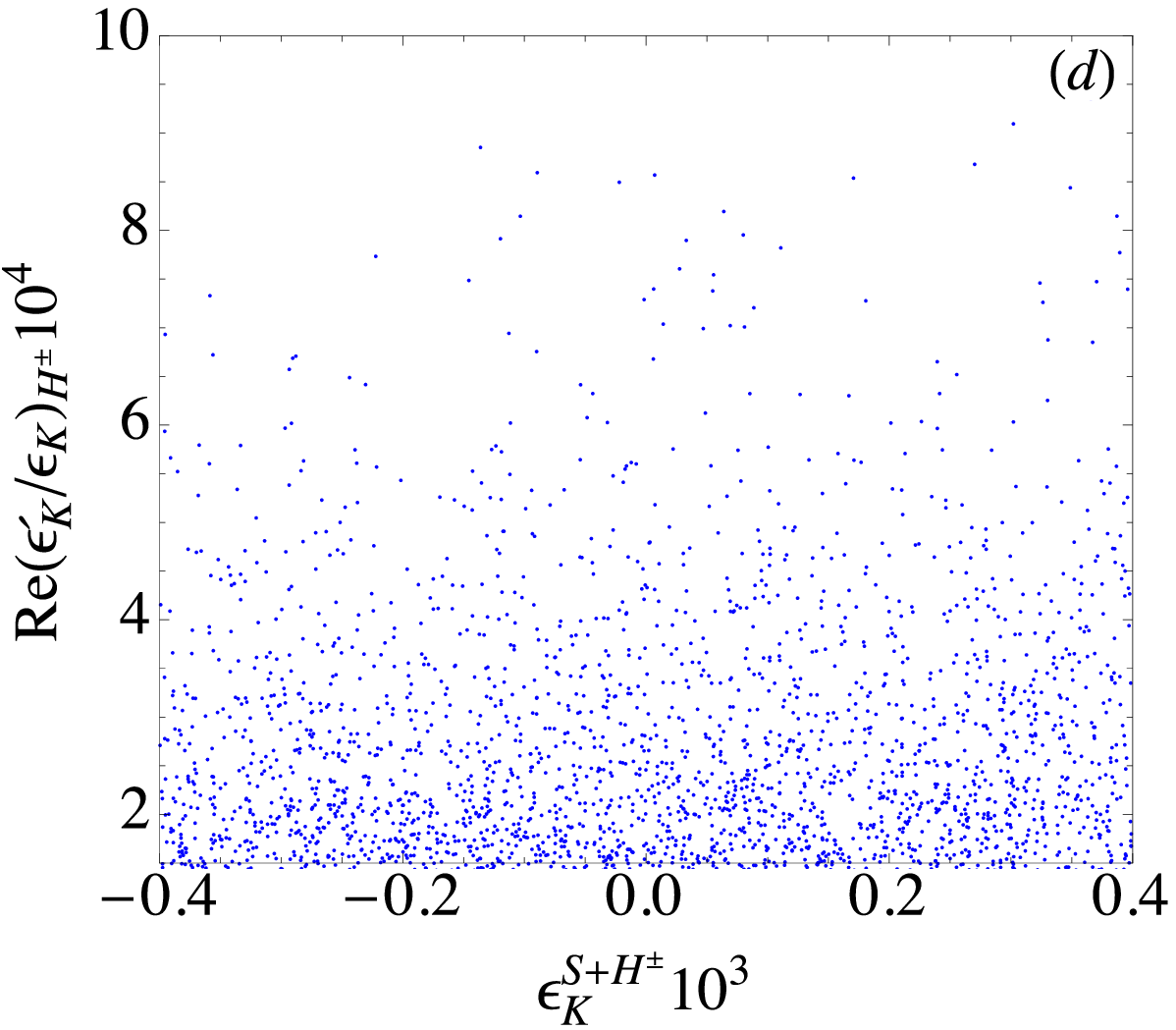}
 \caption{$Re(\epsilon'_K/\epsilon_K)_{H^\pm}$ (in units of $10^{-4}$) as a function of (a) $m_{H^\pm}$, (b) $\chi^u_{ct}$, and (c) $\chi^d_{bs}$, and (d) the correlation between $Re(\epsilon'_K/\epsilon_K)_{H^\pm}$ and $\epsilon^{S+H^\pm}_K$ (in units of $10^{-3}$).  }
\label{fig:epdivep}
\end{figure}

 From our analysis, it was found that to enhance $(\epsilon'_K/\epsilon_K)_{H^\pm}$, $\chi^u_{tt}$ and $\chi^u_{ct}$ prefer to be opposite in sign. In the following numerical analysis, we narrow the scan ranges of $\chi^{u}_{tt}$ and $\chi^u_{ct}$ to be:
    \begin{align}
    0.4 \leq \chi^u_{tt} \leq 0.8\,, \ -0.8 \leq \chi^u_{ct} \leq 0.1\,.
    \end{align}
   In order to include the  tree-induced $\Delta S=2$ effects shown in Eq.~(\ref{eq:HS2}), we set the relevant parameters as:
   \begin{equation}
  | \chi^{d*}_{ds} \chi^d_{sd}| \leq 0.8 \times 10^{-6}\,, ~|\theta_{CP}| \leq \pi\,.
   \end{equation}
 Since we now need to combine the $H^\pm$-induced box diagrams and the $H(A)$-induced tree diagrams for the $\Delta S=2$ process, to satisfy the constraints of $\Delta M_K$ and $\epsilon_K$ and enhance $\epsilon'_K/\epsilon_K$, the taken values of $|\chi^{d*}_{ds}\chi^{d}_{sd}|$ are smaller than those shown in Fig.~\ref{fig:DMKeK_S}. 
 When the constraints from the $B$ and $K$ systems are taken into account, in which the sampling data points for the scan are $2.5\cdot 10^{7}$,  the dependence of $Re(\epsilon'_K/\epsilon_K)_{H^\pm}$ (in units of $10^{-4}$)  on the parameters is given as follows: Fig.~\ref{fig:epdivep}(a) shows $Re(\epsilon'_K/\epsilon_K)_{H^\pm}$ as a function of  $m_{H^\pm}$; Fig.~\ref{fig:epdivep}(b) and (c) are the dependence of $\chi^u_{ct}$ and $\chi^d_{bs}$, respectively, and  Fig.~\ref{fig:epdivep}(d) shows the correlation between $Re(\epsilon'_K/\epsilon_K)_{H^\pm}$ and $\epsilon^{\rm S}_K$ + $\epsilon^{\rm H^\pm}_K$ (in units of $10^{-3}$). Since the dependence of $\chi^d_{bd}$ is similar to that of $\chi^d_{bs}$, we do not show the case for $\chi^d_{bd}$. From these plots, we see that although we cannot push $(\epsilon'_K/\epsilon_K)_{H^\pm}$ up to $O(10^{-3})$, the charged-Higgs effects can lead  to $(\epsilon'_K/\epsilon_K)_{H^\pm} \sim 8 \times 10^{-4}$. 
  
  \subsection{ Charged-Higgs contributions to $K \to \pi \nu \bar \nu$ }
  
  In this subsection, we discuss the charged-Higgs contributions to  the $K^+ \to \pi^+ \nu \bar\nu$ and $K_L \to \pi^0 \nu \bar\nu$ processes. 
 As mentioned before,  the $W^\pm H^\pm$ and $G^\pm H^\pm$ box diagrams  are suppressed by  $m^2_\tau/m^2_W \zeta^\ell_\tau t_\beta$,  where although there is a $t_\beta$ enhancement factor, their contributions are still small and negligible. The Wilson coefficient  from the $H^\pm H^\pm$ box diagram can be enhanced through  the $(\zeta^\ell_\tau t_\beta)^2$ factor; however, its  sign is opposite to that of the SM, so that  it has a destructive effect on the SM results. Thus, we cannot rely on $H^\pm H^\pm$ to enhance $BR(K^+ \to \pi^+ \nu \bar\nu)$ and $BR(K_L \to \pi^0 \nu \bar\nu)$. Hence, the main charged-Higgs effect on the $d \to s \nu \bar\nu$ process is derived from the $Z$-penguin diagram. 
  
\begin{figure}[phtb]
\includegraphics[scale=0.55]{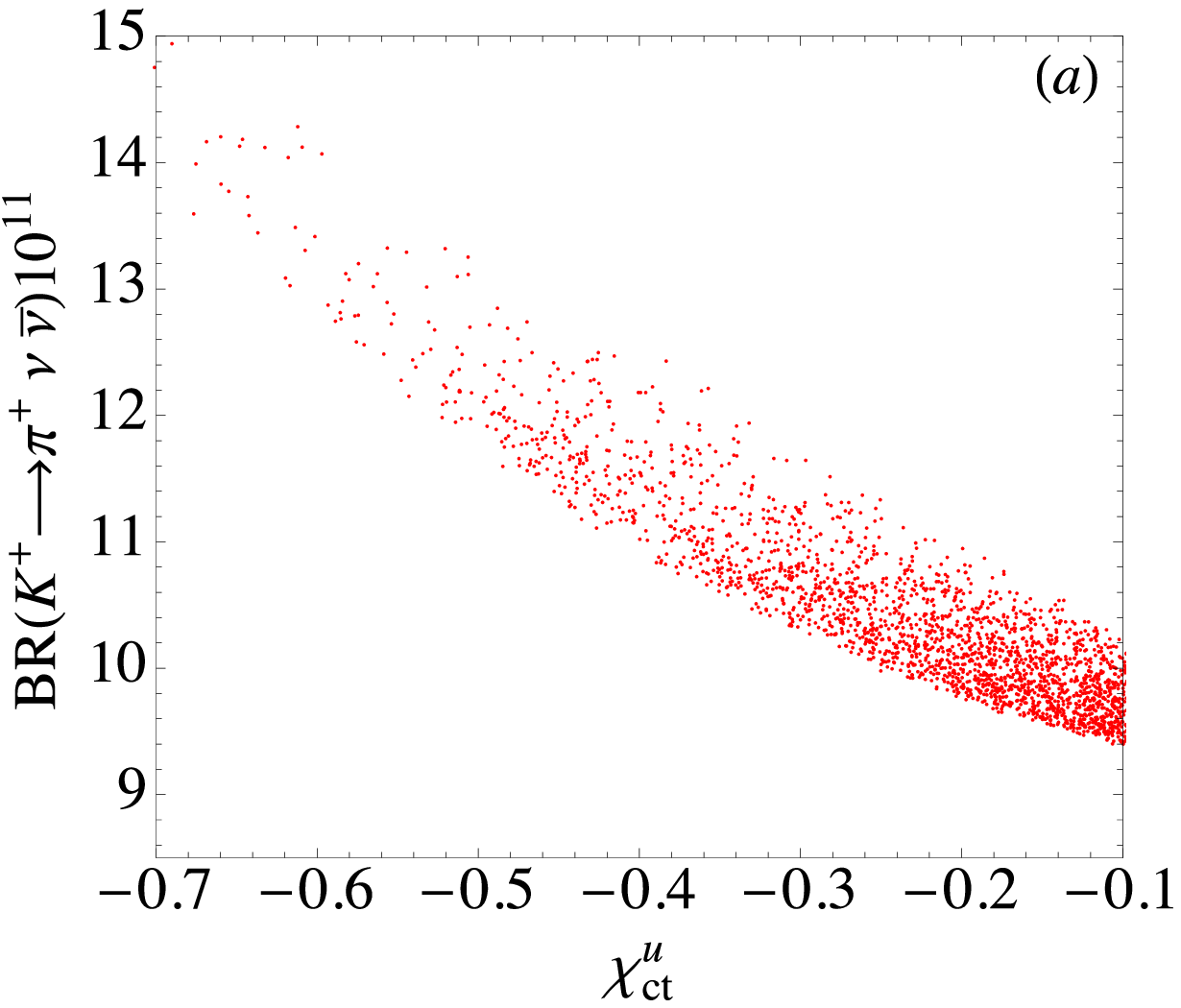}
\includegraphics[scale=0.55]{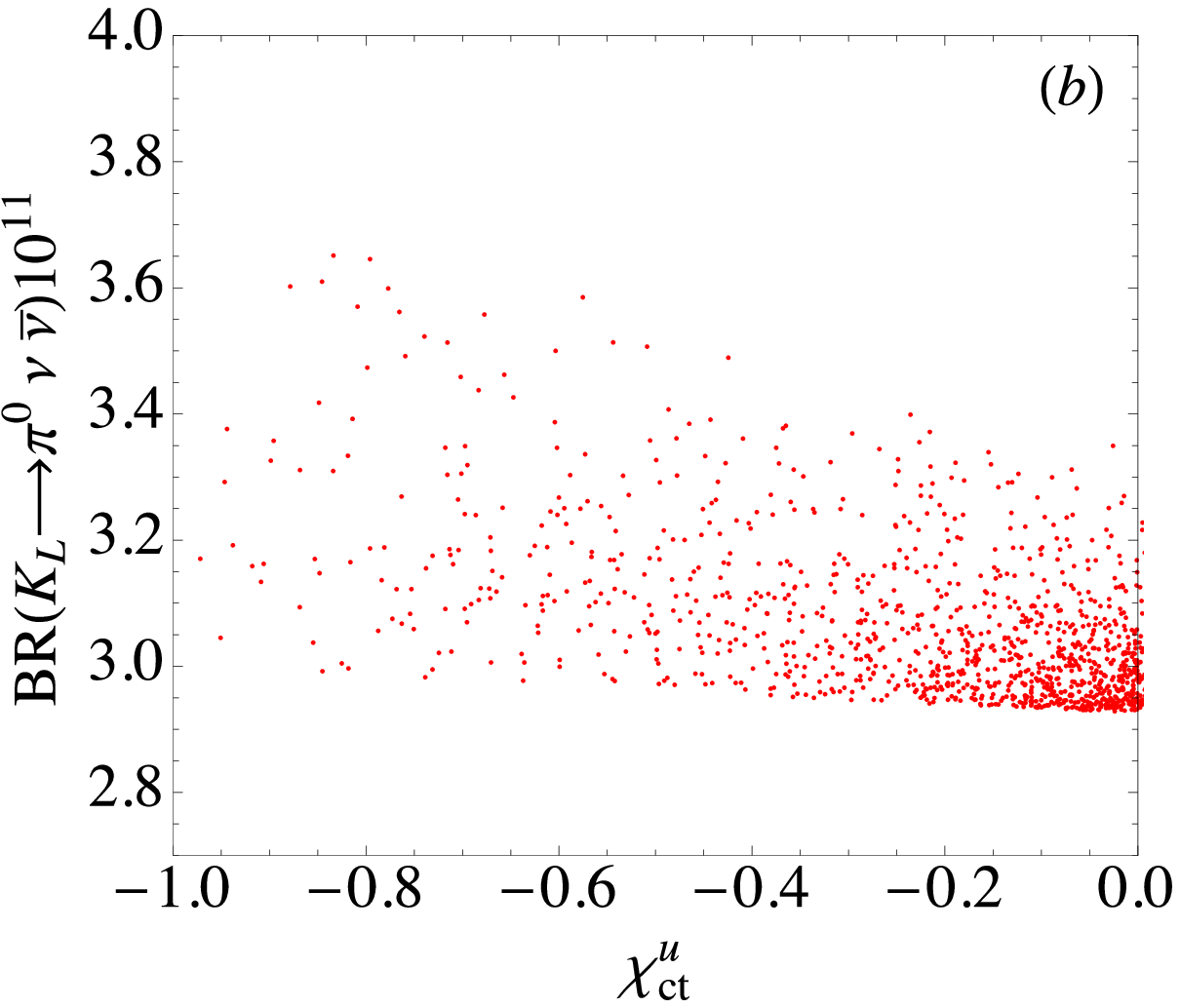}
 \caption{  (a) $BR(K^+ \to \pi^+ \nu \bar\nu)$ and (b) $BR(K_L \to \pi^0 \nu \bar\nu)$ as a function of $\chi^u_{ct}$. }
\label{fig:K_chict}
\end{figure}

\begin{figure}[phtb]
\includegraphics[scale=0.6]{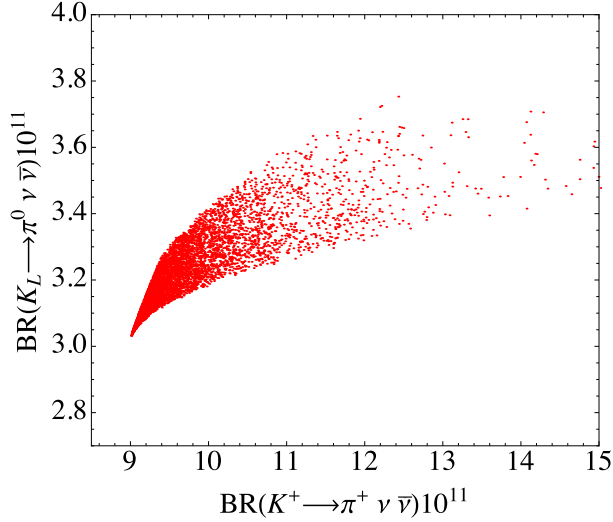}
 \caption{  Correlation between $BR(K^+ \to \pi^+ \nu \bar \nu)$ and $BR(K_L \to \pi^0 \nu \bar\nu)$.  }
\label{fig:Kp_KL}
\end{figure}

\begin{figure}[phtb]
\includegraphics[scale=0.55]{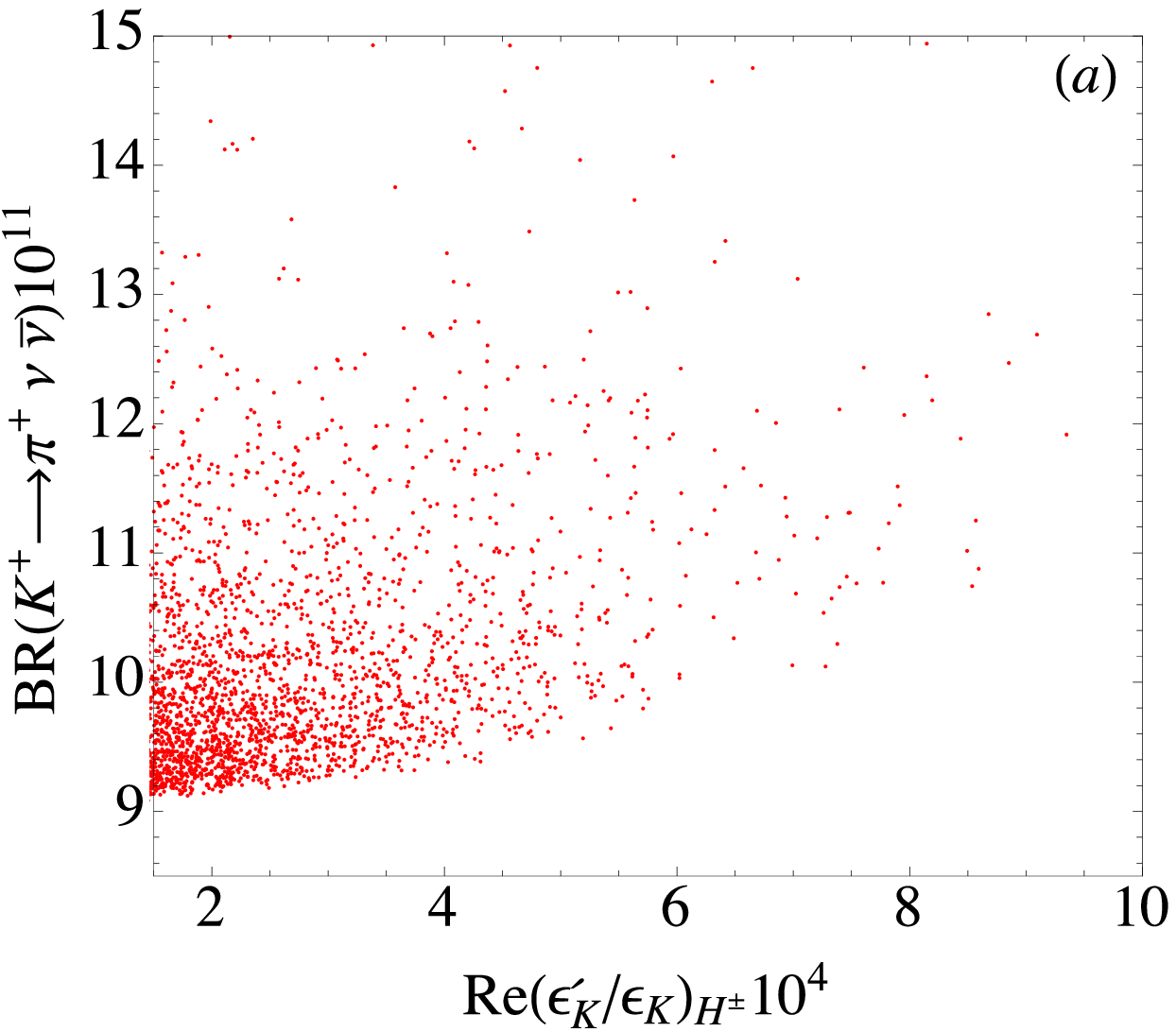}
\includegraphics[scale=0.55]{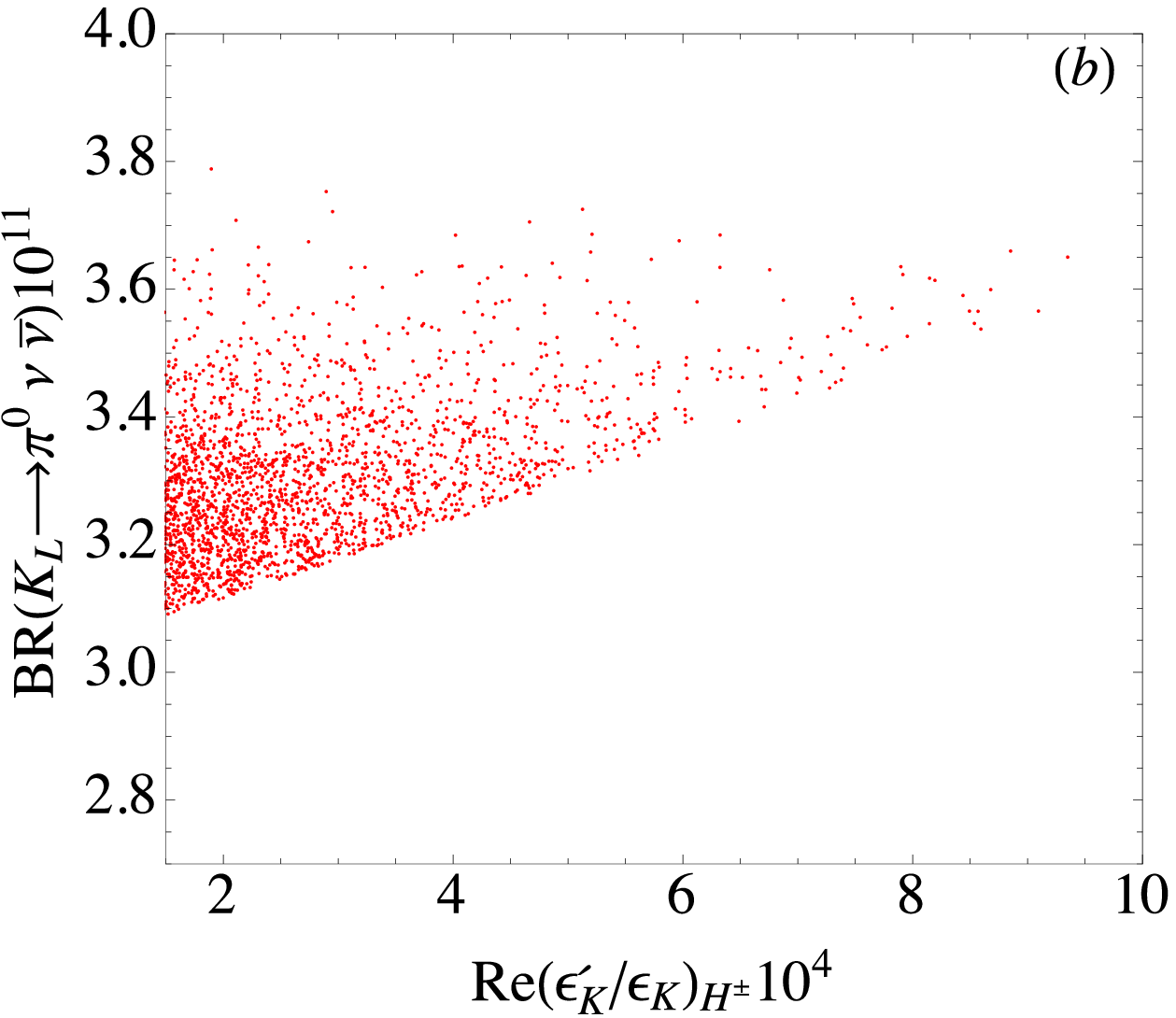}
 \caption{  Correlation between $(\epsilon'_K/\epsilon_K)_{H^\pm}$ and (a) $BR(K^+ \to \pi^+ \nu \bar\nu)$ and (b) $BR(K_L \to \pi^0 \nu \bar\nu)$. }
\label{fig:Kpi_ep}
\end{figure}

  With the input parameter values, the BRs for the $K^+ \to \pi^+ \nu \bar\nu$ and $K_L \to \pi^0 \nu \bar\nu$ processes in the SM can be estimated to be $BR(K^+ \to \pi^+ \nu \bar\nu)\approx 8.8 \times 10^{-11}$ and $BR(K_L \to \pi^0 \nu \bar\nu)\approx 2.9 \times 10^{-11}$. Using the  parameter values, which are constrained by the $B$-meson and $K$-meson data, we calculate the charged-Higgs contributions to $K^+ \to \pi^+ \nu \bar\nu$ and $K_L \to \pi^0 \nu \bar\nu$, where the BRs (in units of $10^{-11}$) as a function of $\chi^u_{ct}$ are shown in Fig.~\ref{fig:K_chict}(a) and (b). In order to suppress the contribution from the $H^\pm H^\pm$ diagram, we take $\chi^\ell_\tau =1$. From the plots, we clearly see that $BR(K^+ \to \pi^+ \nu \bar \nu)$ can be enhanced to $\sim 13\times 10^{11}$ while $BR(K_L \to \pi^0 \nu \bar \nu)$ is enhanced to $\sim 3.6 \times 10^{-11}$. Since the CP violating source in the charged-Higgs loop is the same as that of  the SM, the  $K_L \to \pi^0 \nu \bar\nu$ enhancement is limited. Although the charged-Higgs cannot enhance $K^+\to \pi^+ \nu \bar\nu$ by a factor of 2,  it can increase the SM result by $60\%$.
  For clarity, we show the correlation between  $BR(K^+ \to \pi^+ \nu \bar \nu)$ and $BR(K_L \to \pi^0 \nu \bar\nu)$ in Fig.~(\ref{fig:Kp_KL}).  In addition, the correlations between the rare $K$ decays and $(\epsilon'_K/\epsilon_K)_{H^\pm}$ are also given in Fig.~\ref{fig:Kpi_ep}(a) and (b). 
  
 \section{Conclusion}
  
  We comprehensively studied the $Re(\epsilon'_K/\epsilon_K)$ and the rare $K^+(K_L) \to \pi^+(\pi^0) \nu \bar\nu$ decays in the type-III 2HDM, where the Cheng-Sher ansatz was applied, and the main CP-violation phase was still from the CKM matrix element $V_{td}$ when the Wolfenstein parametrization was taken.  We used $|\Delta M^{\rm NP}_{K}| < 0.2 \Delta M^{\rm exp}_K$ and $|\epsilon^{\rm NP}_K | < 0.4 \times 10^{-3}$ to bound the free parameters. 
  
  The charged-Higgs related parameters, which contribute to $\Delta M_K$ and $\epsilon_K$,  also contribute to $\Delta M_{B_d, B_s}$ and $B\to X_s \gamma$ processes. When the constraints from the $K$ and $B$ systems are satisfied, we found that it is possible to obtain $(\epsilon'_K/\epsilon_K)_{H^\pm} \sim 8 \times 10^{-4}$ in the generic 2HDM, where the dominant effective operator is from the electroweak penguin $Q_8$.

 The dominant contribution to the rare $K\to \pi \nu \bar\nu$ decays in the type-III 2HDM is the  $H^\pm$-loop $Z$-penguin diagram. With the same set of constrained parameters, we found that $K_L \to \pi^0 \nu \bar\nu$ can be slightly enhanced to $BR(K_L \to \pi^0 \nu \bar\nu)\sim 3.6 \times 10^{-11}$, whereas $K^+\to \pi^+ \nu \bar\nu$ can be  enhanced to  $BR(K^+\to \pi^+ \nu \bar\nu)\sim 13 \times 10^{-11}$. Although the BRs of the rare $K\to \pi \nu \bar \nu$ decays cannot be enhanced by one order of magnitude in the type-III 2HDM, the results are still located within the detection level in the KOTO experiment at J-PARC and the NA62 experiment at CERN.
 
\section*{Acknowledgments}

This work was partially supported by the Ministry of Science and Technology of Taiwan,  
under grants MOST-106-2112-M-006-010-MY2 (CHC).

%%%%%%%%%%%%%%%%%%%%%%%%%%%%%%%%%%%%%%%%%%%%%%%%%%%%%%%%%%%%%%%%%%%%%%%%
\bibliographystyle{unsrt}

\end{document}